\documentclass{CSML}
\pdfoutput=1

\usepackage{lastpage}
\newcommand{\dOi}{13(4:12)2017}
\lmcsheading{}{1--\pageref{LastPage}}{}{}%
{Nov.~01,~2016}{Nov.~21, 2017}{}

\usepackage{hyperref}
\hypersetup{hidelinks}

\usepackage{graphicx}
\usepackage{subcaption}
\usepackage{stmaryrd}
\usepackage{MnSymbol}
\usepackage{amsthm}
\usepackage{marginnote}
\usepackage{color}
\usepackage{xcolor}
\usepackage{afterpage}
\usepackage{xspace}
\usepackage{mathpartir,url,needspace}

\usepackage{listings}
\usepackage{macros-ASP}

\newcommand{\intro}[1]{\emph{#1}}

\newcommand{\maos}{multi-active objects\xspace}
\newcommand{\creol}{\textsc{Creol}\xspace}
\newcommand{\jcobox}{\textsc{JCoBox}\xspace}
\newcommand{\absl}{\textsc{ABS}\xspace}
\newcommand{\ambient}{\textsc{AmbientTalk}\xspace}
\newcommand{\encore}{\textsc{Encore}\xspace}
\newcommand{\asp}{\textsc{ASP}\xspace}
\newcommand{\proactive}{\textsc{ProActive}\xspace}
\newcommand{\multiasp}{\textsc{Multi}\asp}

\newcommand{\COG}{\textsc{cog}\xspace}
\newcommand{\COGS}{\textsc{cog}s\xspace}
\newcommand{\COGClass}{\texttt{COG}\xspace}
\newcommand{\threadannot}{\code{@DefineThreadConfig}\xspace}
\newcommand{\mathsizebegin}{\begin{small}\begin{mathpar}}
\newcommand{\mathsizeend}{\end{mathpar}\end{small}}
\newcommand{\Cn}{{\nt{cn}}}

\theoremstyle{plain}
\newtheorem{proposition}{Proposition}[section]
\newtheorem{lemma}[proposition]{Lemma}
\newtheorem{theorem}[proposition]{Theorem}

\DeclareMathOperator{\Method}{Method}

\definecolor{verylightgray}{RGB}{247,247,247}

\begin{document}

%
%


\title{Multi-active Objects and their Applications}

\author[Ludovic Henrio and Justine Rochas]{Ludovic Henrio}	
\address{Universit\'e C\^ote d'Azur, CNRS, I3S, France} 

\author[]{Justine Rochas}
\email{ludovic.henrio@cnrs.fr, justine.rochas@unice.fr} 

\titlecomment{An extended version of this article is published as a research 
report~\cite{HR-LMCSRR-2017}.}


\begin{abstract}
  \noindent In order to tackle the development of concurrent and distributed 
systems, the active object programming model provides a high-level abstraction 
to program concurrent behaviours. There exists already a variety of active 
object frameworks targeted at a large range of application domains: modelling, 
verification, efficient execution. However, among these frameworks, very few  
 consider a multi-threaded execution of active objects. Introducing  
\emph{controlled} parallelism within active objects enables overcoming some of 
their limitations. In this paper, we present a complete framework around the 
multi-active object programming model. We present it through \proactive, the Java 
library that offers multi-active objects, and through \multiasp, the programming 
language that allows the formalisation of our developments. We then show how to  compile 
an active object language with cooperative multi-threading into 
multi-active objects. 
This paper also presents different use cases and the development support to 
illustrate the practical usability of our language. Formalisation of our work 
provides the programmer with guarantees on the behaviour of the multi-active 
object programming model and of the compiler.
\end{abstract}

\maketitle

\section{Introduction}
\label{sec:introduction}

Systems and applications nowadays run in a  concurrent and distributed  manner. In 
general, existing programming models and languages lack   
adequate abstractions for handling the concurrency of parallel programs and for writing  
distributed applications. 
Two well-known categories of concurrency bugs are associated
with synchronisation. On the one hand, deadlocks appear when
there is a circular dependency between tasks. On the other
hand, data races correspond to the fact that several  
threads access a shared resource without proper synchronisation.  In 
object-oriented  programming, allowing several threads of control to execute 
methods breaks  the encapsulation property of objects. Industrial programming 
languages generally do  not enforce by default a safe parallel execution nor 
offer a  programmer-friendly way to program concurrent applications. In Actors and active 
object languages, an application is designed as a set 
of independent entities only communicating by passing messages. This approach makes the 
programming of distributed and concurrent systems easier. 

The active object programming  model~\cite{AOSurvey2017} reconciles object-oriented 
programming and 
concurrency. Each active object has its own  local memory and cannot access the 
local memory of other active objects.  Objects can only exchange 
information through  message passing, implemented with requests sent to active 
objects. This  characteristic makes object accesses easier to trace, and thus 
it is easier to  check their safety. Also, the absence of a shared memory between 
active objects makes  them well-adapted to a distributed execution. 
Section~\ref{sec:background} will give an overview of active-object 
languages. 

Active object models however have limitations in terms of efficiency on 
multicore environments and of deadlocks. Concerning deadlocks, one solution is to allow 
the currently executing thread to be released 
and allow the current active object to handle another request. This approach is called 
cooperative multi-threading and is used in several languages like for example ABS 
(abstract behavioral specification) language~\cite{Johnsen:2010:ACL:2188418.2188430}.

However cooperative multi-threading relies on the programmer's expertise to place release 
points 
which is not always realistic. Additionally, cooperative multi-threading does not solve 
the problem of the 
inefficiency in distributed architectures made of several multicore machines.
This leads us to the 
design of a new model based on active objects: multi-active objects~\cite{Henrio2013}. 
This model enables local multi-threading inside an active object. In  active 
object models,   an activity 
is the unit of composition of the model: an activity is a set of entities evolving 
independently and asynchronously from other objects. In classical active-object models, 
an activity is a set of objects associated with a request queue and a single thread. In 
the multi-active object an 
activity is a single active object equipped with a request queue, it may serve one or 
several requests in parallel. This programming model relies on the notion of 
\emph{compatibility 
between requests}, where two compatible requests can safely be run in parallel. 
\emph{Groups} of methods are introduced to easily express compatibility between requests: 
two requests can run in parallel if they target methods belonging to two compatible 
groups.
\multiasp~\cite{Henrio2013}, is a calculus that formalises the 
multi-active objects.

Concerning the definition of \multiasp, compared to~\cite{Henrio2013}, this article 
presents a new version of the operational 
semantics of the language, and includes an extension that deals with thread 
management;  this article also presents a new debugging tool for multi-active objects.
This 
article is a follow up of the article presented at Coordination 2016~\cite{Henrio2016}. 
Compared to~\cite{Henrio2016}, one contribution of this article is a full overview on 
the 
multi-active programming model including the different programming tools that we provide 
for this programming model. The strongest improvement featured by this article 
compared 
to~\cite{Henrio2016} is the formalisation of the translation and the correctness of the 
translation from ABS programs 
into multi-active objects, and in particular the definition of the relationship between 
\absl\ configurations and their translation in \multiasp. A brief summary of the proof is 
presented in this article and the details of the proof can be found 
in~\cite{HR-LMCSRR-2017}.
From a practical point 
of view, the multi-active object model is distributed as part of the \proactive Java 
library\footnote{\url{https://github.com/scale-proactive}} that supports the distributed 
execution of active objects. The proof allows us to provide more 
 precise conditions of applicability of the equivalence 
between ABS programs and their translation.
We summarise below our main contributions.

\begin{itemize}
 \item We introduce the multi-active object programming model, together with the library 
 that supports it, \proactive, and with its formalisation language, \multiasp. 
 This language is also equipped with advanced features regarding the scheduling of 
 requests: thread management and priority of execution. The operational semantics of 
 \multiasp is presented in two parts: the base semantics of the calculus, and an 
 extension dealing with thread management.
 \item  We show the practical usability of the language in two ways. First, we illustrate 
 the programming model with a case study based on a distributed peer-to-peer system. 
 Second, we show a tool that displays the execution of multi-active objects and that 
 helps 
 in debugging and tuning multi-active object-based applications.
 \item We present an approach for encoding cooperative active objects into multi-active 
 objects. We specify a backend translator and prove  the equivalence between the ABS 
 program and its translation in \multiasp\footnote{The backend is available at: 
 \url{https://bitbucket.org/justinerochas/absfrontend-to-proactive}}.
\end{itemize}

Section~\ref{sec:background} details the background of this paper, the active object 
programming model; it provides a
comparison of active object languages and positions our contribution on multi-active 
objects. 
Section~\ref{sec:framework} introduces our multi-active object framework, its semantics, 
and the semantics of the thread management functionality. 
Section~\ref{sec:encoding} presents the automatic translation of \absl programs 
into \proactive programs, shows the properties of the translation, and provides a 
discussion 
relating this work with the other ABS backends and other related works. 
Section~\ref{sec:conclusion} concludes this article.

\section{Background: Active Objects}
\label{sec:background}
After a brief introduction to active objects and actors, this section provides an 
analysis 
of the characteristics that can be used to classify active-object 
languages. Then \asp, 
ABS, and Encore are presented in more details. These languages have been chosen for the 
following reasons: 
ASP is the language that we extend in our multi-active object model, ABS is the 
language for which we provide a backend in
Section~\ref{sec:encoding}, and Encore features a few novel features including some 
controlled parallelism inside active objects.
We refer  to~\cite{AOSurvey2017} for a  
precise description of other active object models including \jcobox, AmbientTalk, and 
Creol. The section concludes by a focus on the related works featuring some form of 
multi-threaded active objects.
 
\subsection{Origins and Context}
\label{sbsc:background-origins}

The active object programming model, introduced 
in~\cite{Lavender:1996:AOO:231958.232967}, has one global objective: facilitate 
the correct programming of concurrent entities. The active object paradigm 
derives from actors~\cite{Agha:1986:AMC:7929}. Actors are concurrent entities 
that communicate by posting messages to each other. Each actor has a mailbox 
and a thread that processes the messages of the mailbox. Only the processing of 
a message can have a side effect on the actor's state. Once a message is posted 
to an actor, the sender continues its execution, without knowing when the 
message will be processed. This way the interaction between actors is 
asynchronous, allowing the different entities to perform tasks in parallel.

Active objects are the object-oriented descendants of actors. Like an actor, each active 
object has an 
associated thread and we call \emph{activity} a thread together 
with the objects that are accessible by this thread. 
Active objects communicate 
using asynchronous method invocations:
when an object invokes a 
method of an active object, this creates a \emph{request} that is 
posted in a mailbox (also called \emph{request queue})
on the callee side. On the invoker side, the execution continues while the request is 
being processed. On the callee side, the request  
is dropped in 
the request queue and waits until its turn comes to be 
executed. 

Like in object-oriented programming, the invocations of methods on 
active objects can return a result. Since these invocations are asynchronous, their 
result 
cannot be known just after the invocation. To represent the 
expected result and to allow the invoker to continue its execution asynchronously, a 
placeholder is created for the result of the request. This place holder is called a 
\emph{future}, which is a promise of response that will be later filled by the result of 
the 
request. A future is \emph{resolved} when the value of the future is computed 
and available. 
Futures have seen their early days in  
MultiLisp~\cite{Halstead:1985:MLC:4472.4478} and  
ABCL/1~\cite{Yonezawa:1986:OCP:28697.28722}. They have been  formalised 
in ABCL/f~\cite{Taura94abcl/f:a}, in~\cite{Flanagan:1999:SFA:968699.968700}, 
and, more recently, in a concurrent lambda 
calculus~\cite{Niehren:2006:CLC:1226601.1226607}, and in the Creol 
language~\cite{deBoer:2007:CGF:1762174.1762205}.  In summary, active objects 
and actors enforce decoupling of activities: each activity has its own memory 
space, manipulated by its own thread. This strict isolation of concurrent 
entities makes them locally safe and also suited to distributed systems.

\subsection{Design of Active Object Languages}
\label{sbsc:background-implementations}

Existing implementations of the active object programming model make different 
design choices for four main characteristics that we have extracted and listed 
below, as answers to design questions. 

\medskip\noindent\textit{How are objects associated to activities?}
\label{sec:activity}
We distinguish three different ways to map objects to threads/activities in 
active object languages: 

\begin{description}
 \item[\begin{small}Uniform object model\end{small}] all objects are active objects
  with their own execution
  thread. All communications between them necessarily create requests. This model is simpler to formalise and reason about, but leads to scalability issues in practice.
  
Creol~\cite{Johnsen:2006:CTO:1226608.1226611, 
	deBoer:2007:CGF:1762174.1762205} and Rebeca~\cite{DBLP:conf/csicc/SirjaniMM01}  are 
	uniform active object languages. \item[\begin{small}Non uniform object 
	model\end{small}] some objects are not
  active objects; they are called \emph{passive} objects and are only accessible by one 
  active
  object.
  This model is scalable
  as it requires less communication and
  less threads, but it is trickier to formalise and reason about because some of the 
  objects are only locally reachable and an additional mechanism is necessary to transmit 
   them between active objects.
  Reducing the number of activities also reduces the number of
   globally accessible references in the system, and thus enables the
  instantiation of a large number of objects.
Non-uniform active objects
reflect better the design of efficient distributed applications where many objects 
are created but only some of them are remotely accessible.

ASP~\cite{Caromel:2005:TDO:1952047}, AmbientTalk~\cite{Dedecker:2006:APA:2171327.2171349, 
	Cutsem:2007:4396972}, and Joelle~\cite{Clarke:2008:MOA}  are typical 
non-uniform active-object 
languages.  
Orleans~\cite{BernsteinB16} is a  recent industrial active object language with a 
non-uniform active object model.
 \item[\begin{small}Object group model\end{small}]
	an activity is made of a set of objects sharing 
	an execution thread, but all
	objects can be invoked from another activity. This
	approach has a good scalability and propensity to formalisation, but
	the addressing of all objects in distributed settings is difficult to maintain 
	because it requires a distributed referencing system able to handle a large number of 
	objects.
	
	ABS~\cite{Johnsen:2010:ACL:2188418.2188430} and 
	\jcobox~\cite{Schafer:2010:JGA:1883978.1883996} have concurrent object groups but  
	\jcobox additionally allows data sharing for immutable objects. in \jcobox, the 
	object 
	groups are called coboxes, while they are called \COGS (concurrent object groups) in 
	ABS.
\end{description}
\needspace{2ex}

\medskip\noindent\textit{How are requests scheduled?}
We distinguish three request scheduling models:
\begin{description}
 \item[\begin{small}Mono-threaded scheduling\end{small}]
	within an active object, requests are executed sequentially without
	interleaving. This model is simple to reason about and has some strong 
	properties but is the most prone to deadlock and potential inefficiency because one 
	activity is blocked as soon as a synchronisation is required.
	
	ASP~\cite{Caromel:2005:TDO:1952047}, 
	AmbientTalk~\cite{Dedecker:2006:APA:2171327.2171349, 
		Cutsem:2007:4396972}, and Rebeca~\cite{DBLP:conf/csicc/SirjaniMM01} 
	feature Mono-threaded 
	scheduling but Rebeca and AmbientTalk perform no synchronisation  and have no
	deadlock.
	
\begin{figure}
	\centering
	\begin{subfigure}{.53\textwidth}
		\centering
		\includegraphics[height=3.6cm]{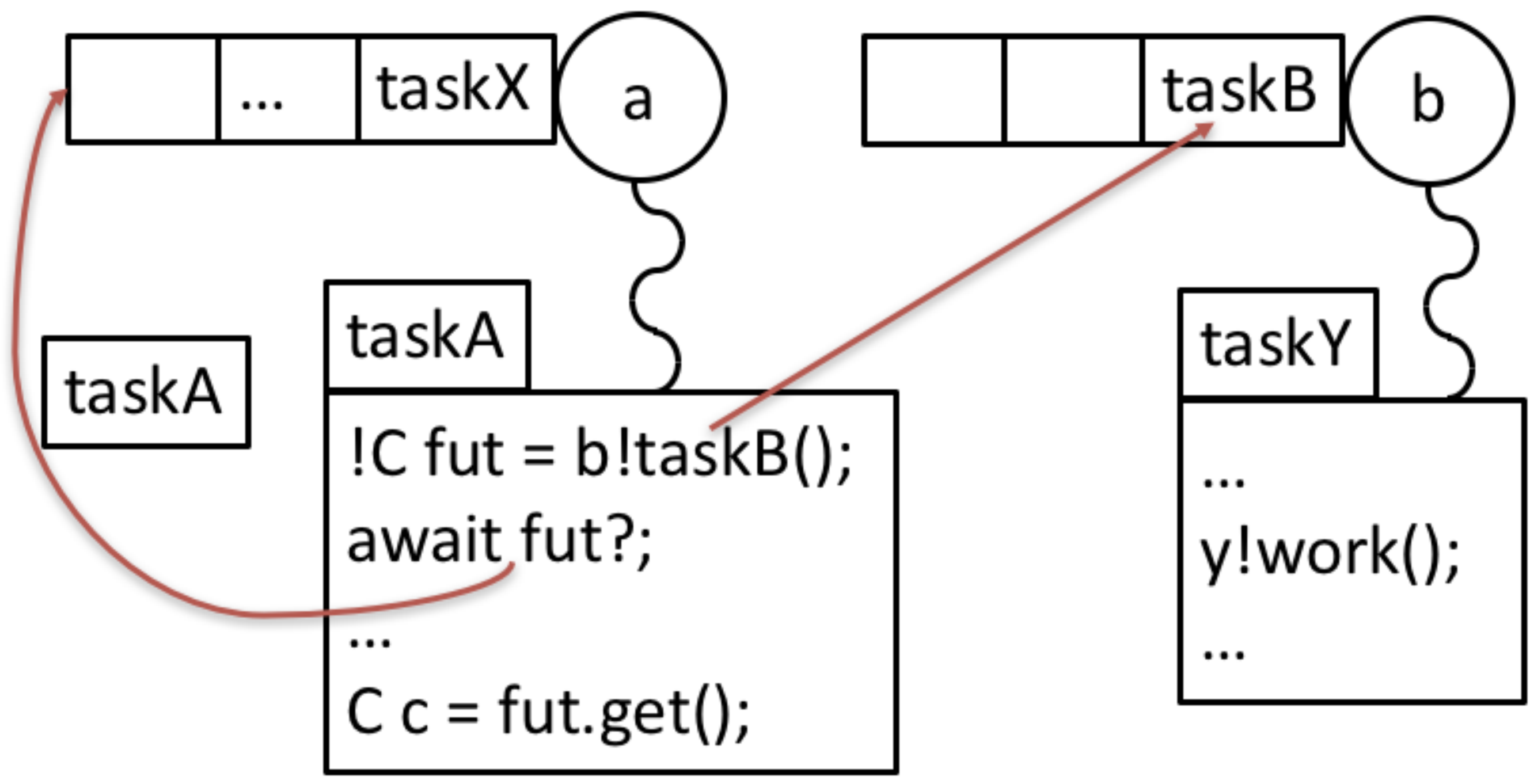}
		\caption{Configuration 1.}
		\label{fig:background-creol-1}
	\end{subfigure}%
	\begin{subfigure}{.47\textwidth}
		\centering
		\includegraphics[height=3.6cm]{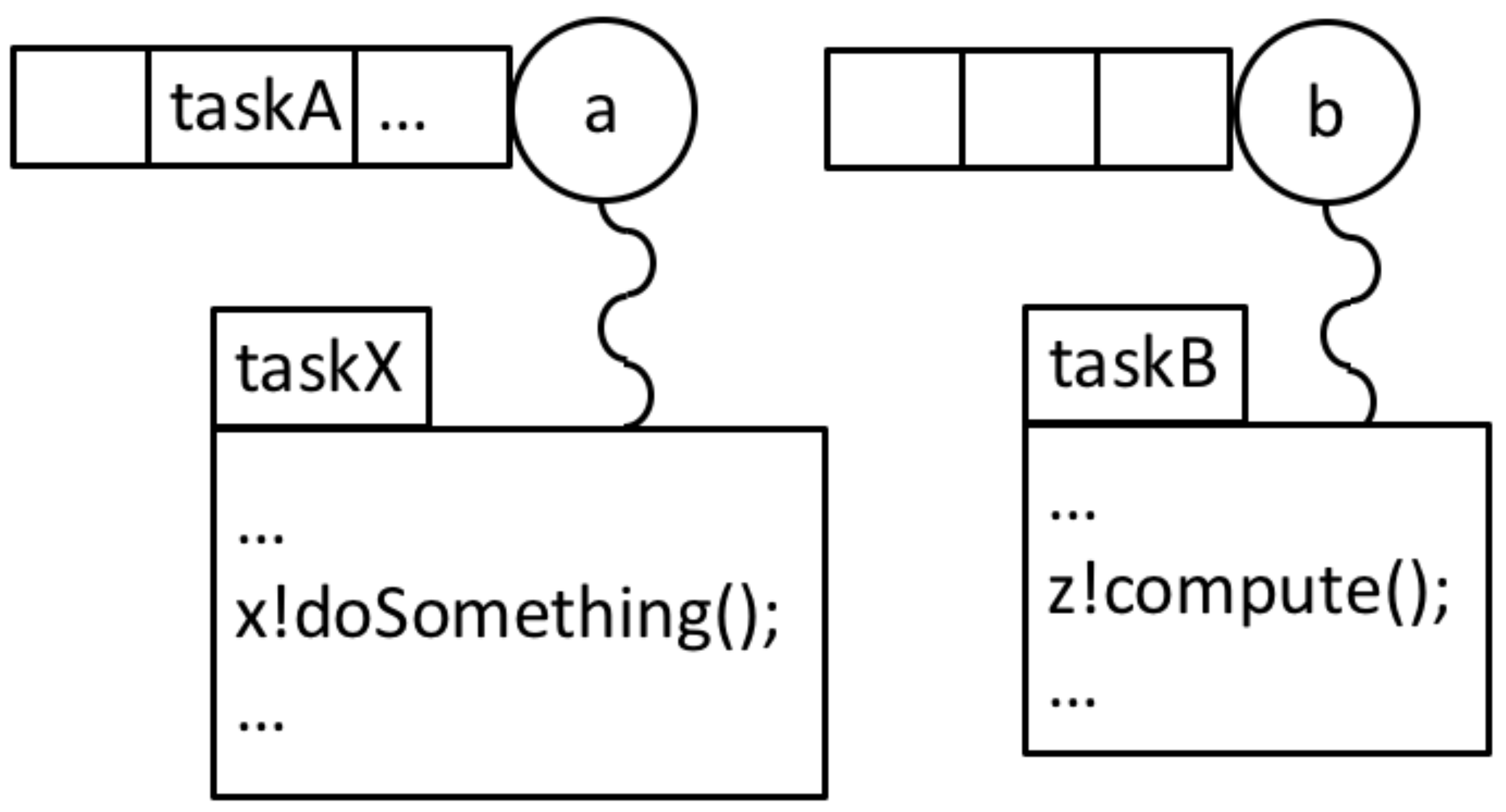}
		\caption{Configuration 2.}
		\label{fig:background-creol-2}
	\end{subfigure}
	\caption{Cooperative scheduling in Creol.}
	\label{fig:background-creol}
\end{figure}
	 \item[\begin{small}Cooperative scheduling\end{small}]
	a running request can explicitly release the execution thread to let
	another request progress. Requests are not processed in parallel but they
	might interleave. While no data races are 
	possible, the status of an object might be difficult to predict when a request is 
	restored after a release point if the cooperatively scheduled requests modify the 
	object state.
	
Creol~\cite{Johnsen:2006:CTO:1226608.1226611, 
		deBoer:2007:CGF:1762174.1762205}, 
		\jcobox~\cite{Schafer:2010:JGA:1883978.1883996}, 
		and ABS~\cite{Johnsen:2010:ACL:2188418.2188430} are typical examples of 
		cooperative scheduling languages.
	Figure~\ref{fig:background-creol} illustrates cooperative scheduling, based on a 
	Creol example. It shows two objects \code{a} and \code{b} each with a queue of 
	awaiting tasks (rectangles). The currently executed task is shown below the object.
	In 
	Figure~\ref{fig:background-creol-1}, object 
	\code{a} does an asynchronous method invocation on object \code{b}, and then 
	awaits for the result.	
	As the future \code{fut} is not yet resolved  when the \code{await} statement 
	executes, this suspends the execution of the current task \code{TaskA} that returns 
	to the pending task queue. At this point, another request for this object can 
	start or resume; in the example a new active task is \code{TaskX} is started. 
	Figure~\ref{fig:background-creol-2} shows a later 
	configuration, after several steps of execution;  
	\code{taskA} is still in the queue of object \code{a}, \code{taskX} is running. In 
	object \code{b}, \code{TaskY} finished and the next task has started. 
	Finally, when \code{TaskB} and \code{taskX} finish, \code{taskA} can resume 
	and retrieve the value computed by \code{TaskB} with a blocking call to 
	\code{.get()}, shown at the end 	of \code{taskA} in 
	Figure~\ref{fig:background-creol-1}.

 \item[\begin{small}Multi-threaded scheduling\end{small}] several threads are running in 
 parallel in the same actor or active object. Either several requests can be served in 
 parallel, or a data-level parallelism allows a 
 request to be processed by several threads. In this context, an \emph{activity} becomes 
 a set 
 of threads together with the objects they manipulate: there is one activity per active 
 object or per object group. Consequently, data races are 
 possible but only within an activity. An additional mechanism is 
	necessary to control which threads can run in parallel. Parallelism provides 
	efficiency 
	at the expense of possible data-races. Somehow, similarly to cooperative 
	multithreading, efficiency and expressiveness are gained at the expense of some 
	possible incoherency in the object state. However, the two approaches offer different 
	trade-offs: in 
	cooperative multithreading incoherences can be limited by removing release points, 
	while in multi-threaded scheduling incoherences can be limited by controlling which 
	threads can run in parallel. 
On the other hand, multithreading enables parallelism internally to an active object	
which 
is more efficient in some cases like in a multicore setting.

	\multiasp and parallel actor monitors~\cite{Scholliers:2014:PAM} feature 
	multi-threaded scheduling where several 
	requests can be served in 
	parallel by the same active object. Multi-threaded actors~\cite{AzadbakhtBS-ICE16} 
	feature multi-threading, but with each thread hosted by a different active-object (an 
	actor contains several active objects in the terminology of~\cite{AzadbakhtBS-ICE16}).
	Section~\ref{sec:RW-MAO} discusses in more details multithreading in active objects 
	and actors.
\end{description}


\medskip\noindent\textit{How much is the programmer aware of  
asynchronous aspects?}
Some active object languages use a specific syntax for asynchronous method 
calls and a specific type for futures.  This makes the programmer aware of 
where  synchronisation occurs in the program. When the asynchronous 
invocation is explicit, there often exists a special type for future objects, 
and an operator to access the future's value. On the contrary, asynchronous 
method calls and futures can also be implicit: this enables transparency of 
distributed aspects, and facilitates the transmission of future references 
(sometimes called first-class futures~\cite{Henrio:2010:FCF:2031978.2032019}). In this 
case, 
there is almost no syntactic difference between a distributed program with asynchronous 
invocations and usual objects, consequently writing simple distributed programs is 
easier. On the 
other hand, explicit manipulation of 
futures allows the programmer to better control the program execution, but  
also requires a better programming expertise. 

	Creol~\cite{Johnsen:2006:CTO:1226608.1226611, 
	deBoer:2007:CGF:1762174.1762205}, \jcobox~\cite{Schafer:2010:JGA:1883978.1883996}, 
 ABS~\cite{Johnsen:2010:ACL:2188418.2188430}, and Encore~\cite{Brandauer2015} have 
 explicit future types and explicit asynchronous invocations contrarily to 
 ASP~\cite{Caromel:2005:TDO:1952047} where there is no specific type for futures or 
 active 
 objects.
 
 \medskip\noindent\textit{How are handled the results of asynchronous method 
 invocations?}
 What distinguishes active objects from actors is 
 the fact that communication between activities is performed by invocation of methods,  
 according to the object's interface, and not through send and receive operations.  
This distinction in the terminology is also highlighted in  Akka~\cite{AkkaBook} where 
``actors'' are distinguished from 
``typed 
 actors'' that are in fact active objects. The different languages propose 
 different ways to 
 handle method results:
 \begin{description}
 	\item[No return value] In some languages the methods of an active object cannot 
 	return any value. This is for example the case of Rebeca or Scala 
 	actors~\cite{Haller:ScalaActors-2009}.
 	\item[Futures] The most classical approach is to use a future to represent the 
 	expected result of an asynchronous method call. When the returned value is needed, 
 	the program is interrupted until the value becomes available. The futures can be 
 	explicitly visible to the programmer like in ABS or transparent like in ASP. In the 
 	second case there is no need for a specific instruction to wait for the future, the 
 	current thread is automatically blocked when the value associated to a future is 
 	\emph{needed}. This mechanism is called \emph{wait-by-necessity}. In languages 
 	with cooperative scheduling, it is possible to let another request execute while the 
 	future is awaited (e.g. using the \code{await} statement in ABS).
 	\item[Asynchronous futures] Some languages use a future to represent the 
 	expected result of an asynchronous method call but provide no synchronisation on 
 	futures. Instead a continuation is triggered upon the resolution of the future. This 
 	continuation can be more or less explicit depending on the language. Akka and 
 	AmbientTalk feature asynchronous futures.  For example, in AmbientTalk,
 	 a 
 	future access creates and asynchronous 
 	invocation that will be triggered after the future has been resolved. The 
 	event-based execution of the different activities makes 
 	sequences of actions more difficult to enforce. However, an AmbientTalk program is 
 	always partitioned into separate event handlers that maintain their own 
 	execution context, making the inversion of control easier to handle.
 \end{description}

\subsection{Abstract Behavioural Specification.}\label{sec:ABS}
\begin{figure}[t]
	\begin{center}
		\begin{small}
			$         \begin{array}{r@{}c@{}l@{}r@{}}
			g&::=& b \bnfor x? \bnfor g \land g'       &\text{guard}\\
			s&::=& \eskip\bnfor x\!=\!z\bnfor\texttt{suspend}\bnfor
			\texttt{await}\ g\bnfor\ereturn e \bnfor\eif ess  \bnfor s\!\semi\! s  
			&\text{statement} \\
			z&::=&  e \bnfor e.\name m(\vect e)\bnfor e!\name
			m(\vect e) \bnfor \enew{[\texttt{local}] \C(\vect e)}\bnfor 
			x.\text{get}&\text{\hspace{-13.5ex}expression 
				with 
				side effect}\\
			e&::=& v\bnfor x\bnfor\ethis\bnfor\text{\it arithmetic-bool-exp}   
			&\text{expression}\\
			v&::=& \enull\bnfor \text{\it primitive-val} &\text{value}
			\end{array}$
		\end{small}
	\end{center}
	\caption{Syntax of the concurrent object layer of ABS (definition of method, class, 
	and types omitted). }\label{fig:classABS}
\end{figure}
\begin{figure}
	\begin{minipage}{0.5\textwidth}
		\begin{lstlisting}[label=lst:background-abs, caption=ABS program code.]
		BankAccount ba = new BankAccount();
		Transaction t = new local Transaction(ba); 
		WarningAgent wa = new WarningAgent();
		Fut<Balance> bfut = ba!apply(t);
		await bfut?;
		Balance b = bfut.get;
		wa!checkForAlerts(b);
		b.commit() \end{lstlisting}
	\end{minipage}
	\begin{minipage}{0.49\textwidth}
		\centering
		\includegraphics[scale=0.25]{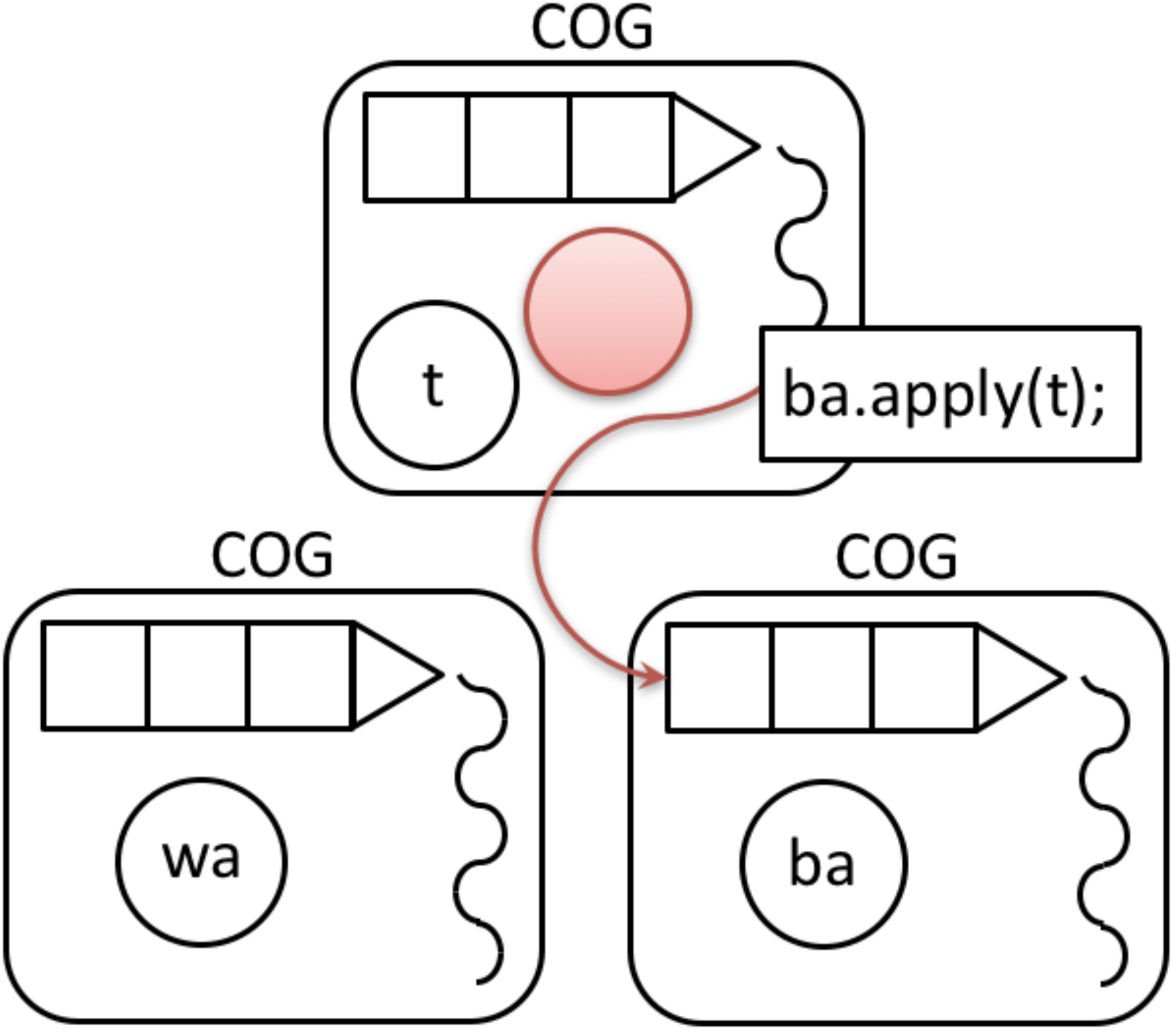}
		\caption{ABS program execution.}
		\label{fig:background-abs}
	\end{minipage}
\end{figure}
\textit{ABS}~\cite{Johnsen:2010:ACL:2188418.2188430} is an 
object-oriented modelling language based on active objects. \absl takes its 
inspiration in \creol for the cooperative scheduling and in \jcobox for the 
object group model. It  uses explicit asynchronous method calls and 
futures. \absl is intended for modelling and verification of distributed 
applications. The object group model of \absl is represented by Concurrent 
Object Groups (\COGS). A \COG manages a request queue and a set of tasks that 
have been created as a result of asynchronous method calls to any of the 
objects inside the \COG. Inside a \COG, only one task
 is active at any time. New objects can be 
instantiated in a new \COG with the \code{new} keyword. In order to instantiate 
an object in the current \COG, \code{new local} must be used instead. In the example, the 
transaction \code{t} is instantiated in the  \COG\ that is running the current (main) 
method.  Contrarily to 
\jcobox, \absl makes no difference on the object kind: all objects can be 
referenced from any \COG and all objects can be invoked either synchronously or 
asynchronously. Listing~\ref{lst:background-abs} shows an \absl program that 
creates two new \COGS and performs asynchronous method calls.  Note the specific 
syntax (\code{!}) for asynchronous method  invocation and the \code{await} instruction on 
Line~5 that releases the current task if the future stored in \code{bfut} is not yet 
available.
Figure~\ref{fig:background-abs} pictures the sending of the \code{apply} 
request to the remote \COG. In the illustration, \COGS are large rectangles with round 
angles, request queues are depicted at the top of \COGS, and objects are symbolised by 
circles.
The concurrent object part of the syntax of ABS is shown in Figure~\ref{fig:classABS}. 
In this syntax, x range over variable names, the overline notation ($\vect{e}$) is used 
for lists, \textit{arithmetic-bool-exp} range over arithmetic and boolean expressions, 
and \textit{primitive-val} stands for primitive (integer and boolean) values. The most 
significant statements and constructs have been described in the example. 
 Note that field access is 
restricted to the current object ($\ethis$).

\absl comes with numerous engines (see
\url{http://abs-models.org/}) for verification of concurrent and distributed 
applications: a deadlock analyser~\cite{Giachino2015}, a resource consumption 
analyser~\cite{Johnsen201567}, a termination and cost analyser 
COSTABS~\cite{Albert:2012:CCT:2103746.2103774,Albert2014}, and a 
 verifier for generic program properties
KeY-ABS~\cite{Din:August:2015,Din:November:2015}. In addition to verification 
tools, \absl tools also comprise a frontend compiler and several backends that 
translate \absl programs into  Maude, Java, or 
Haskell~\cite{Bezirgiannis2016}. This work is partially based on the Java backend.

\subsection{Asynchronous Sequential Processes.} ASP~\cite{Caromel:2004:ADO:964001.964012} 
is an active object language 
with a non uniform 
active object model; it is intended to be close to  realistic implementations of 
distributed systems. 
  ASP features \emph{mono-threaded scheduling} and futures are 
implicitly created from asynchronous remote method calls. \asp features a 
wait-by-necessity behaviour upon access to an unresolved 
future. \asp has proven determinism properties and formalises object 
 groups and software
components~\cite{Caromel:2005:TDO:1952047}. Communications ensure causal ordering of 
requests.
\proactive \cite{Baduel2006} is the Java 
library that enforces the 
\asp semantics. As the active object model of \asp is transparent, 
\proactive active objects follow as much as possible the syntax of standard Java. The 
only 
syntactic difference is the
 \code{newActive} primitive that is used to create an active object. 
Listing~\ref{lst:background-proactive} 
shows the same example as in the \absl above, written in \proactive. 
Nevertheless, the semantics of the two programs is different since in \absl 
the future is always resolved when the \code{checkForAlerts} request is sent.

\begin{lstlisting}[float, label=lst:background-proactive, caption=An example of ProActive program. \emph{node} is not defined here., emph={node}, emphstyle=\itshape]
BankAccount ba = PAActiveObject.newActive(BankAccount.class, null, node) ;
Transaction t = new Transaction(ba); 
WarningAgent wa = PAActiveObject.newActive(WarningAgent.class, null, node);
Balance b = ba.apply(t); // t is deeply copied
wa.checkForAlerts(b); // if b is a future it is passed transparently
b.commit(); // a wait-by-necessity is possible here
\end{lstlisting}

\proactive is intended for distribution, it forms a complete middleware 
that supports the deployment of applications on distributed infrastructures such 
as clusters, grids, and clouds. To this end, when an active object is created, it 
is registered in the Java RMI registry, RMI being the main communication layer 
used in \proactive. Consequently, passive objects are deeply copied when 
communicated between activities, the different copies are then handled independently and 
can be in an inconsistent state. The advantage of this approach is its  scalability 
and its coherency with the mechanism of RMI, and consequently its practical effectiveness.
Because \proactive is  a Java API, it must be integrated with the standard 
Java programming language. Proxies are created to  represent generalised  pointers 
to futures and active objects; consequently futures and active object cannot be of 
primitive type and  a method that returns a primitive type would be called 
synchronously.

\subsection{Encore}
Encore~\cite{Brandauer2015}  has an active object 
programming model mixed with other parallel patterns. \encore is essentially 
based on a non-uniform object model but uses capacities to handle the concurrent access 
to passive objects, thus enabling shared references to passive objects. \encore uses 
cooperative scheduling of requests 
that is similar to \creol and \absl.  Futures are explicitly typed and their value must 
be 
 retrieved via a \code{get} construct.  \encore natively features parallel constructs 
 other 
than asynchronous method invocations. Internal parallelism can be spawned explicitly 
 through \code{async} blocks, or implicitly through parallel 
combinators~\cite{party-coordination}.  \encore unifies all its parallel constructs with 
the 
use of futures for 
handling asynchrony. A chaining operator \code{$\rightsquigarrow$} can add a 
callback to a future, in order to execute it when the future is resolved. Thus Encore 
features both synchronisation on futures and asynchronous futures. Cooperative 
scheduling 
and future 
chaining mix explicit and automatic synchronisation, and save the 
programmer from the burden of precisely placing all the release points in the 
program.

\subsection{Multi-threaded Actors and Active Objects}~\label{sec:RW-MAO}
In this paper, we focus on 
multi-threaded active objects~\cite{Henrio2013}, that are characterised by a controlled 
parallelism within an activity. Contrarily to Encore, parallelism does not occur 
inside a request, but between requests, and under the control of the programmer since only 
requests tagged as compatible can run in parallel. Comparing multi-threaded scheduling to cooperative scheduling is more complex as the concurrency model is much different: on one hand 
cooperative scheduling prevents race conditions that might occur within multi-threaded 
active objects (e.g. if the wrong requests are declared compatible), but on the other hand 
multi-threaded active objects provide a parallel execution that is more efficient and allow 
the programmer to control which requests  run in parallel. Section~\ref{sec:encoding} 
 shows how cooperative scheduling can be faithfully encoded with multi-threaded active 
objects and  provides a  deeper technical insight on the comparison between 
the two approaches.

In~\cite{Hayduk2013} the authors enable automatic parallel execution of requests inside 
actors. They use transactional memory to undo local side-effects that may be conflicting. 
However this work supports actors interacting only by asynchronous messages; it does not 
take into account any other form of synchronisation, like futures in 
active objects.

Other  works introduced multi-threading inside actors or active objects. 
Parallel actors monitors~\cite{Scholliers:2014:PAM} (PAM) was designed at the same time 
as \multiasp; it provides an interface to schedule the parallel service of requests in an 
actor. The framework is richer and allows the programmer to express any scheduling of 
request and fully control the request treatment. On the contrary, multi-active objects 
only allow the programmer to state which requests  can be served in parallel, possibly 
depending on dynamic criteria. Multi-active objects feature a higher level of abstraction 
where the programmer is unable to reorder request or cancel some of them, overall 
multi-active objects preserve more guarantees of the original actor and active-object 
programming model. In fact, compatibility annotations could be used to generate a 
specific PAM scheduler that would simulate the behaviour of multi-active 
objects.
More recently, \emph{multi-threaded actors} have been proposed~\cite{AzadbakhtBS-ICE16}. 
This approach is quite different from multi-active objects 
or PAM because each active object is  mono-threaded but several active objects
share the same request queue. The parallel treatment of requests in the same request 
queue is  similar to the approach presented in this paper, but the fact that those 
requests are served by different actors makes a significant difference. On one side the 
approach 
of \cite{AzadbakhtBS-ICE16} ensures the absence of race-conditions but on the other side, 
it gives no guarantee on which object will serve which request. 
This approach is well adapted to stateless objects, but not adapted to stateful objects, 
as different requests can be handled by different objects. Indeed 
consider two requests targeting a multi-threaded actor, if the first one modifies the 
state of one of the actor's object and the second request is handled by a different 
object, then the state modification will not be visible by the second request.


\section{The Multiactive Object Framework}
\label{sec:framework}
This section presents the Multiactive object programming model, its formal model called 
\multiasp, and its implementation and support inside the \proactive library.
Even if the works were realised independently, the concurrency and annotation system of  
multi-threaded active objects 
shares some similarities with JAC (Java annotations for concurrency)~\cite{CPE:CPE956}.

\subsection{Programming Model and Language}
\label{sbsc:framework-language}

\subsubsection{Principles}

Multi-active objects enable local parallelism inside active objects. To this end, the 
programmer can annotate the class of an 
active object with information about concurrency by defining a compatibility 
relationship between requests. The principle is to allow two independent requests 
to execute in parallel, 
but prevent requests that could conflict on some resources to run concurrently.

In the RMI style of programming, every remote invocation to an object will be 
run in parallel with no synchronisation. As a result, data-races can happen on 
concurrently accessed resources. A classic approach to solve this problem is to 
protect concurrent executions with a lock, but this approach is too 
fine-grained to be scalable, and possibly error-prone. By nature, active 
objects materialise a much safer model where no inner concurrency is possible. 
However, we aim at a  programming model that is locally concurrent and more flexible than 
the mono-threaded active objects, but more constrained and less error-prone 
than bare synchronisation of threads. The principle relies on the notion of 
\emph{compatible requests}. Only requests that are declared compatible can run in 
parallel. To start serving a new request, a multi-active object first checks that it is 
compatible with all the currently served requests, and with the requests that should be 
served before\footnote{Request service in ASP and \proactive follows a FIFO policy.}.

Multi-active objects extend active objects 
by assigning each method to at most one group. These groups define a 
compatibility relationship between requests: only the 
requests that target methods belonging
to compatible groups can be executed in parallel, whereas the others will be 
guaranteed not to run concurrently. Since the groups and their 
compatibilities are defined with annotations, the application logic is not 
mixed with concurrency features. Two groups should be made compatible if 
their methods do not access the same data, and if the methods of the two groups can be 
executed in any order. Compatibility is a commutative, non-transitive relation. A group 
may be declared to be compatible with itself.
We base our multi-active object framework on active objects \`a la \asp. 
An 
active object can be turned into a multi-active object by applying the following 
design methodology:

\begin{itemize}
 \item Without annotations, a multi-active object behaves identically to a mono-threaded 
 active object,
no race condition is possible, but no local parallelism is possible either.  
Methods assigned to no group are incompatible with all the methods: by default methods 
are 
conflicting, and no data-race is possible.
 \item If  some  parallelism  is  desired,  for  efficiency  reasons  or  because 
 of potential  deadlocks, each remotely invocable method can be assigned to a group. Then 
 compatibility between the groups can be defined based on the fields
accessed by each method.  
 \item If even more parallelism is desired, the programmer has two non-exclusive options: 
  protect the access to the fields by a locking mechanism
and declare more groups as compatible, or define a compatibility function that decides at 
runtime which requests are compatible, depending on the request parameters or on the 
object state.
\end{itemize}

In all cases, we assume that the programmer defines the groups and their 
compatibilities correctly. Dynamic checks or static analysis should be added 
to ensure that no race condition appear at runtime, but this is 
out of scope of this paper.

Multi-active objects provide a customisable trade-off between gaining some parallelism at 
the intra-object level, removing some deadlocks, and loosing some safety in terms of 
absence of data races (compared to classical active objects where the absence 
of data races is guaranteed). Compatibilities must be defined carefully to prevent safety 
problems. 

\subsubsection{Multi-active objects in Practice}

\proactive offers an implementation of the multi-active object programming model 
as an extension of its active object implementation. A set of  Java 
annotations can be used for the specification of multi-active object notions. 
The annotations are processed at runtime, which enables  to 
decide dynamically on the compatibility of requests. The three main annotations are the 
following:

\begin{itemize}
	\item A \code{@Group} annotation can be declared on top of a class to define a group 
	of requests. Request in the same group share the same compatibility requirements. 
	\item A \code{@MemberOf} annotation can be defined on top of a method definition, and specifies the group to which the method belongs.
	\item A \code{@Compatible} annotation can be used to declare the groups that are compatible, i.e. to declare the requests that can be run in parallel safely. 
\end{itemize}

\begin{lstlisting}[float, label=lst:application-peer, caption=The Peer class of the fault tolerant broadcast application.]
@DefineGroups({
	@Group(name="gettersOnImmutable", selfCompatible=true),
	@Group(name="dataManagement", selfCompatible=true),
	@Group(name="broadcasting", selfCompatible=false),
	@Group(name="monitoring", selfCompatible=true)
})
@DefineRules({
	@Compatible({"gettersOnImmutable", "broadcasting", "dataManagement"}),
	@Compatible({"gettersOnImmutable", "monitoring"})
})
public class Peer implements Serializable, RunActive { 
	private LongWrapper identifier;
	private Zone zone;

	@MemberOf("gettersOnImmutable")
	public LongWrapper getIdentifier() { ... } 

	public BooleanWrapper join(Peer p, int dimension) { ... }

	@MemberOf("dataManagement")
	public AddResponse add(AddQuery query) { ... }

	@MemberOf("broadcasting")
	public void broadcast(Key constraint, RoutingPair message) { ... }
}
\end{lstlisting}

As an example, consider a distributed peer-to-peer system implemented with 
multi-active objects.  The purpose of this application is to offer a high 
performance distributed system for data storage. Each peer is represented by a 
multi-active object that is instantiated from a \code{Peer} class. The 
\code{Peer} class is partially shown in Listing~\ref{lst:application-peer}.  A 
peer has operations that are related to the structure of the peer-to-peer 
system, as well as operations that are related to data management. In the annotation 
\code{DefineGroup}, 
four groups are defined (Lines 1 to 6). They correspond to the different concerns 
addressed by the class.
All the groups except \code{broadcasting} are self compatible: they allow several 
requests 
of the same group to be run in parallel.
 Then compatibility rules are defined between those groups (in the annotation 
 \code{DefineRules}), for 
example Line 8 declares that the methods of groups \code{gettersOnImmutable}, 
\code{broadcasting}, and \code{dataManagement} can be executed in parallel. One can 
notice that the \code{join} 
method does not belong to any group: a \code{join} request will 
always be executed alone, it is incompatible with all other requests. On the contrary 
requests on method \code{getIdentifier} can be executed in parallel with any other 
request.

\subsubsection{\multiasp}

\newcommand{\WITH}{\text{\underline{with}}}
\renewcommand{\iff}{\text{\,\underline{iff}\,}}
\newcommand{\absred}{\stackrel{\textsc{a}}{\to}}
\newcommand{\absredN}{\stackrel{\textsc{a}}{\to}^*} \newcommand\redN{\to^*}
\newcommand\Compatible{\nt compatible} \newcommand\Limit[1]{\mathcal{L}_{#1}}
\newcommand\NbActive[1]{\text{Active}({#1})} \newcommand\Rc{\approx}

\begin{figure}[hbt]\vspace{-2ex}
	\mysyntax{\mathmode}{
\simpleentry P  [] \vect{C}\ \wrap{\vect{x}\ s} 
[program]\\                                                              
\simpleentry S  [] \name m(\vect{x})                                   [method signature]\\
\simpleentry C  [] \dsclass{C}(\vect{x}){\ \wrap{\vect{x}\ \vect{M}}} \qquad~    [class]\\
\simpleentry M  [] S\wrap{\vect{x}\ s}                                                               [method definition]\\
\simpleentry s []
      \eskip\bnfor x=z\bnfor \ereturn e \bnfor s\semi s [statement]\\
\simpleentry z [] e \bnfor e.\name m(\vect e) \bnfor\enew{C(\vect e)}\bnfor 
\newActive{C(\vect e)}\qquad[expression with side effects]\\
\simpleentry e [] v\bnfor x\bnfor\ethis\bnfor\text{\it arithmetic-bool-exp}   [expression]\\
\simpleentry v [] \enull\bnfor \text{\it primitive-val} [value]
	}
	\caption{The class-based static syntax of MultiASP.}
	\label{fig:background-masp-static-syntax}
\end{figure}

\multiasp is the active object programming language that extends \asp for the 
support of multi-active objects. \multiasp formalises the multi-active objects 
that are implemented in \proactive, and allows us to reason on the execution of 
\proactive programs. A seminal version of \multiasp is given 
in~\cite{Henrio2013}. This paper shows an updated version  based on 
classes instead of object instances. 
 In this paper, we also extend the operational semantics with advanced scheduling 
 capabilities. \multiasp is an imperative 
programming language, whose syntax is inspired from object-oriented core 
languages resembling to Featherweight 
Java~\cite{Igarashi:2001:FJM:503502.503505}. As can be seen in 
Figure~\ref{fig:background-masp-static-syntax}, the syntax of \multiasp is also 
 close to  \proactive programs.  A program 
consists of a set of classes and one main method. Classes, methods, and 
statements are standard. In the syntax, $x$ ranges over variable names, 
$C$  over class names, and $\m$  over method names. We characterise 
a list of elements with the overlined notation. The list $\vect{x}$ denotes 
local variables when it appears in method bodies and  object fields  in class 
declarations.  In \multiasp, as in \proactive, there are 
two ways to create an object: \code{new} creates a new object in the current 
activity (a passive object), and \code{newActive} creates a new active object. 
Also, no syntactic distinction exists between local and remote invocations, 
$e.\name m(\vect e)$ is the generic method invocation, that triggers an asynchronous 
method invocation if 
the targeted object is active or a local synchronous method invocation if it is passive. 
Similarly, as synchronisation on 
futures is  transparent and handled through wait-by-necessity, there is no 
particular syntax for interacting with a future. 
Variables in ASP refer either to a local variable of the current method or to a field of 
the current object.
A special variable, $\ethis$, 
enables access to the current object. The sequence operator is associative 
with a neutral skip element: a sequence of instructions can always be rewritten as 
$s;s'$, with $s$ not a sequence.

\newcommand{\Sq}{\nt{F}} \newcommand{\Rq}{\nt{Rq}} \newcommand{\Val}{\nt{val}}
\newcommand{\Sassoc}{\!\shortleftarrow\!}

\begin{figure}[tb]
	\centering 
	\begin{small}
		$\begin{array}{rcl} 
		 \Cn  &::= & \many{\nt{elem} }\\
		\nt{elem}&::= & \fut(f,v,\sigma) \sep  \fut(f,\undefined) \sep  
		\act(\vact,o,\sigma,p,Rq) \\
	 		\nt{v} & ::= & o \sep \vact \sep  \enull\bnfor \text{\it primitive-val}
		\\
		\nt{Storable} & ::= & {[\manyMap{x\mapsto v}]} \sep v \sep f\\
		s & ::= & x=\bullet \sep \eskip\bnfor x=z\bnfor \ereturn e \bnfor s\semi s 		
		\\
		 z & ::= & e \bnfor e.\name m(\vect e) \bnfor\enew{C(\vect e)}\bnfor 
		\newActive{C(\vect e)}		\\

		e & ::= &  v\bnfor x\bnfor\ethis\bnfor\text{\it arithmetic-bool-exp}
		\end{array}$\qquad
		$\begin{array}{rcl}
		\nt{E} & ::= & \{\ell \mid s\} 
\\ 
				\Sq & ::= & E \sep E::\Sq 
		\\ q & ::= & (f,\m,\many{v})\\
		\nt{p} & ::= & \manyMap{q \mapsto\Sq}		\\
		\nt{Rq} & ::= & \emptyset \sep q::\nt{Rq}
		\\ 
		\sigma &::=& \manyMap{o\mapsto\nt{Storable}} \\
		\ell & ::= &  \ethis\mapsto v,\manyMap{x\mapsto v} 
 
		\end{array}$
	\end{small}
	\caption{The runtime syntax ($e$ and $z$ similar to the static 
	syntax).}
	\label{fig:background-masp-runtime-syntax}
\end{figure}

The runtime syntax of \multiasp is shown in 
Figure~\ref{fig:background-masp-runtime-syntax}. The set of elements of a 
\multiasp configuration $\Cn$ are of two kinds: activities and future binders. We 
rely on  
three infinite sets: \emph{object 
locations} in the local store, ranged over by \{$o$, $o'$, 
$\ldots$\}; \emph{active objects names}, ranged over by \{$\alpha$, 
$\beta$, $\ldots$\}; and \emph{future names}, ranged over by \{$f$, 
$f'$, $\ldots$\}. Additionally, the following terms are defined:

\begin{itemize}
	\item There are two kinds of runtime values. \emph{Simple values} ($v$) can be 
	either  
	values of the static syntax ($\enull$, {\it primitive-val}), the location of an 
	object in the local store ($o$), or active object names ($\alpha$). 
	\emph{Storable 
		values} ($\nt{Storable}$) are 
	either objects, futures, or simple values. 
	An object is  a mapping\footnote{We denote  mappings by 
	$\manyMap{\_\mapsto\_}$, 
		and use  union $\cup$ (resp. disjoint union $\uplus$) over mappings. Mapping 
		updates are written $\sigma[x\mapsto v]$, updating the value associated to $x$ in 
		$\sigma$. $dom$ is the domain of a mapping. Additionally, for any vectors 
		$\many y$ and $\many v$ of the same length, $[\manyMap{y\mapsto v}]$ maps the 
		elements of one vector to the other.} 
		from field 
	names to their values: $[\manyMap{x\mapsto v}]$. 
	\item A local environment $\ell$ mapping  local variables (including 
	$\ethis$) to simple values.
	\item A thread $F$ is a stack of methods being executed, 
	where each method execution $E$ consists of a local environment 
	and  a statement $s$ to execute. The first method of the stack is actually 
	executing, the others have been put in the stack due to  
	local synchronous method calls. A special statement $x=\bullet$ allows us to 
	remember the current execution point when a new element is added to the stack. 
	\item Activities are of the form $\act(\vact,o,\sigma,p,\Rq)$. An activity contains 
	terms that define:\\
		-- $\vact$, the \emph{name of the activity}.\\
		-- $o$, the location of the \emph{active object} in $\sigma$.\\
		-- $\sigma$, a \emph{local store} mapping object locations to storable values.
		\\
		-- $\Rq$, a FIFO \emph{queue of requests}, awaiting to be served.\\
		-- $p$, a set of \emph{requests currently served}: a mapping, associating to each 
		currently served request the
		corresponding thread  $\Sq$. 
	\item Future binders are of two forms. The form $\fut(f,\undefined)$ means that the 
	value of the future has not been computed yet: it is  \emph{unresolved}. The form 
	$\fut(f,v,\sigma)$ is used when the value of the future has been \emph{resolved}. 
	The future value can be either a primitive value or a reference to an active or 
	passive object.
	If 	the 	future value references a passive object, then the piece of 
	store $\sigma$ defines the content of this object. As only active objects are 
	remotely accessible, the part of the store 
	referenced by this location must also be transmitted when the future's value is sent 
	back to the caller. This step involves a serialisation mechanism  explained  
	below. Note that the store $\sigma$ can contain references to other futures.
\end{itemize}

An object location  is 
fresh if it does not exist in the store where it is added. A 
future or an activity name is fresh if it does not exist in the  
configuration, we use the following auxiliary functions:

\begin{figure}[bt]
	\centering 
	$\begin{array}{r@{}c@{}ll@{}} 
	\feval{\nt{primitive-val}}{(\sigma+ \ell)}&\triangleq & \nt{primitive-val} &
		\qquad\qquad\qquad\feval{\enull}{(\sigma+ \ell)}\triangleq  \enull\\
	\feval{\vact}{(\sigma+ \ell)}&\triangleq&  \vact 
\\
	\feval{e\oplus e'}{(\sigma+ \ell)}&\triangleq&\feval{e}{(\sigma+ \ell)}\oplus 
	\feval{ e'}{(\sigma+ \ell)} & \text{if 
		$\feval{ e}{(\sigma+ \ell)}$ and $\feval{ e'}{(\sigma+ \ell)}$ are primitive 
		values} \\
	\feval{x}{(\sigma+ \ell)}&\triangleq & \feval{\ell(x)}{(\sigma+ \ell)} & \text{if 
		$x\in\dom(\ell)$}\\
	\feval{x}{(\sigma+ \ell)}&\triangleq& \feval{\ell({\ethis})(x)}{(\sigma+ \ell)} & 
	\text{if $x\notin\dom(\ell)$}\\
	\feval{o}{(\sigma+ \ell)}&\triangleq & o & \text{if }\sigma(o) = f \,|\, 
	{[\manyMap{x\mapsto 
			v}]} \\
	\feval{o}{(\sigma+ \ell)}&\triangleq & \feval{\sigma(o)}{(\sigma+ \ell)} & \text{if 
	}\sigma(o) = o'\,|\, \alpha \,|\, \enull \,|\, \nt{primitive-val}
	\end{array}$
	\caption{Evaluation function}
	\label{fig:translation-evaluation-function}
\end{figure}
\begin{itemize}
	 	\item $\fields{\C}$ returns the fields as defined in the  declaration of the 
	 	class named	 	$\C$. 
 	\item $\bind$ instantiates and initializes a method execution. If $o$ is of class $C$ 
 	and $M$ is a method of $C$ with the signature
 	$m(\overline{y})$ and body 
 	$\{\overline{x}~ s\}$, then:\\
$\bind(o,\m,\many{v'}) = \{ \manyMap{y \mapsto v'},
\manyMap{x \mapsto \enull},\ethis\mapsto o \mid s\}$

	\item $\feval{e}{(\sigma+ \ell)}$ returns the value 
of $e$ by computing the arithmetic and boolean 
expressions and retrieving the values  stored
in  $\sigma$ or  $\ell$, this evaluation function is defined in 
Figure~\ref{fig:translation-evaluation-function}. $\oplus$ stands for any  
arithmetic or binary binary operator (unary operators can be expressed similarly). If one 
member of 
an arithmetic expression is an unresolved future, the function is undefined.
$\feval{e}{(\sigma+ \ell)}$  returns a value and, if the value is 
a
reference to a location in the store, it follows references
recursively; it only returns a location if it points to an object\footnote{The way the 
store is built  guarantees that expression 
evaluation
   terminates if the store is finite.}.
  $\feval{\vect{e}}{(\sigma+\ell)}$ returns the tuple of values of $\vect{e}$.
	\item $\ready$ is a predicate that decides whether a request $q$ is
  ready to be served. We will use $\ready(q,p,\Rq)$ where $p$ the requests currently 
  served  and $\Rq$ the requests that have been received before $q$; it is 
  \emph{true} if $q$ is
  compatible with all requests in $p$  and
   in $\Rq$:
$\ready(q,p,\Rq)=\left(\forall q' \!\in\!\left(\dom(p)\cup 
\nt{Rq}\right).\Compatible(q,q')\right)$.

In the semantics, compatibility is expressed between requests. This is 
expressive enough to encode the compatibility relations that can be written in 
\proactive, including compatibility 
depending on method parameters. It is also easy to extend it in order to define 
compatibility depending on object state. Compatibility on request can be derived from 
group compatibility by stating that two requests are compatible if they target methods 
belonging to compatible groups.
	\item Serialisation reflects the communication 
happening in Java RMI. All references to passive objects
are serialised when communicated between activities, so that they
are always handled locally. We formalise a  serialisation
algorithm that marks and copies the objects to serialise recursively.
The function $\serialise$ takes a value $v$ and a store $\sigma$ and returns a sub-store 
of $\sigma$ containing the serialisation of $v$. It is defined  as the 
mapping verifying 
the following (co-inductive)
constraints:
\mathsizebegin
\begin{array}[t]{@{}l@{}}
\serialise(o, \sigma) = (o\mapsto\sigma(o))
 \cup \serialise(\sigma(o), \sigma)
\qquad\quad
\serialise([\manyMap{x\mapsto v}], \sigma) =
 \bigcup_{v'\in\many{v}} \serialise(v', \sigma)
\\ 
\serialise(f, \sigma) \!=\! \serialise(\alpha, \sigma)\!=\!\serialise(\enull,
\sigma)\!=\! 
\serialise(\text{\it primitive-val}, \sigma) \!=\! \emptyset
\end{array}
\mathsizeend
	\item The function $\rename{\sigma}{\many{v},\sigma'}$ renames the object locations 
	appearing in
$\sigma'$ and $\many{v}$, making them disjoint from the object locations of
$\sigma$; it returns the renamed set of values $\many{v'}$ and another store $\sigma''$, 
as a pair of the form 
$(\many{v'},\sigma'')$.
\end{itemize}

\afterpage{\clearpage}
\begin{figure}[tp]
	\centering 
	\renewcommand{\arraystretch}{0.7} 
	\begin{mathpar}
		\begin{footnotesize}
  \inferrule[Serve]
	{
\ready(q,p,\Rq)\\
q=(f,\m,\many{v}) \\\\
  \bind(o,\m,\many{v})
  = \{\ell\mid s\}
}
{
	\act(\alpha,o,\sigma,p,\nt{Rq}::q::\nt{Rq'})
	\\\\\to
	\act(\alpha,o,\sigma,\{q\mapsto \{\ell|s\}\}\uplus p,\nt{Rq}::\nt{Rq'})}
	
\inferrule[Assign-Local]{
	x \in \dom(\ell)  \\  v=\feval{e}{(\sigma+ \ell)}
	}{
	\act(\alpha,o,\sigma,\{q\mapsto\{\ell \mid x=e;s\}::\Sq\}\uplus p,\nt{Rq}) 
	\\\\\to {\act(\alpha,o,\sigma,\{q\mapsto\{\ell[x\mapsto v] \mid
          s\}::\Sq\}\uplus p,\nt{Rq})}}
          
\inferrule[Assign-Field]{ 
	 \ell(\ethis)=o\\x \!\in\! \dom(\sigma(o)) \\ x\!\notin\!dom(\ell) \\
        \sigma'=\sigma[o\mapsto (\sigma(o)[x\mapsto \feval{e}{(\sigma+\ell)}])] 
       }
	{\act(\alpha,o_\alpha,\sigma,\{q\mapsto\{\ell \mid x=e;s\}::\Sq\}\uplus p,\Rq)
	\to {\act(\alpha,o_\alpha,\sigma',\{q\mapsto\{\ell\mid s\}::\Sq\}\uplus p,\Rq)} }
	
\inferrule[New-Object]{
       \fields{\C}=\many{y}\\
	o~ \fresh\\
	\feval{\many{e}}{(\sigma+\ell)}=\many{v} \\
       \sigma'=\sigma\cup\{o\mapsto[\manyMap{y\mapsto v}]\}      
	}{
	\act(\alpha,o_\alpha,\sigma,\{q\mapsto\{\ell \mid 
	x=\enew{\C(\many{e})};s\}::\Sq\}\uplus p,\Rq)	 \to
	\act(\alpha,o_\alpha,\sigma',\{q\mapsto\{\ell \mid x=o;s\}::\Sq\}\uplus p,\Rq)}
	
\inferrule[New-Active]{
\fields{\C}=\many{y}\\ 
       o,\beta~\fresh  \\
       \feval{\many{e}}{(\sigma+\ell)}=\many{v}\\
       \sigma'=\{o \mapsto[\manyMap{y\mapsto v}]\}\cup\serialise(\many{v},\sigma)
	}{
		\act(\alpha,o_\alpha,\sigma,\{q\mapsto\{\ell \mid x=\newActive{\C(\many{e})};s\}::\Sq\}\uplus p,\Rq)
	\\ \to
	\act(\alpha,o_\alpha,\sigma,\{q\mapsto\{\ell \mid x=\beta;s\}::\Sq\}\uplus p,\Rq)~~ 
	\act(\beta,o,\sigma',\emptyset,\emptyset)
       }
       
\inferrule[Invk-Active]{
  \feval{e}{(\sigma+\ell)}\!=\!\beta \\ 
  \feval{\many{e}}{(\sigma+\ell)}\!=\!\many{v}\\
  f,o~\fresh\\\\ \sigma_1\!=\!\sigma\cup\{o\!\mapsto\!\! f\}
  \\
  (\many{v_r},\sigma_r)\!=\!\rename{\sigma'}{\many{v},\serialise(\many{v},\sigma)}\\
  \sigma''\!=\!\sigma'\cup\sigma_r
}
{
  \act(\alpha,o_\alpha,\sigma,\{q\mapsto\{\ell \mid x=e.\m(\many{e});s\}::\Sq\}\uplus p,\Rq) ~ \act(\beta,o_\beta,\sigma',p',\Rq')\\
\to \act(\alpha,o_\alpha,\sigma_1,\{q\mapsto\{\ell \mid x=o;s\}::\Sq\}\uplus p,\Rq) ~~
\act(\beta,o_\beta,\sigma'',p',\Rq'::(f,m,\many{v_r})) ~~\fut(f,\undefined)
}
    
    \inferrule[Invk-Active-Self]{
	\feval{e}{(\sigma+\ell)}\!=\!\alpha \qquad
	\feval{\many{e}}{(\sigma+\ell)}\!=\!\many{v}\qquad
	f,o~ \fresh\qquad \sigma_1\!=\!\sigma\!\uplus\!\{o\!\mapsto\! f\}
	\\
	(\many{v_r},\sigma_r)=\rename{\sigma_1}{\serialise(\many{v},\sigma)}\\
	\sigma'\!=\!\sigma_r\cup\sigma_1 } {
	\act(\alpha,o_\alpha,\sigma,\{q \!\mapsto\!\{\ell \!\mid\! 
	x=e.\m(\many{e});s\}::\Sq\} 
	\uplus p,\Rq) 
	\\\to \act(\alpha,o_\alpha,\sigma',\{q\mapsto\{\ell \mid 
	x\!=\!o;s\}\!::\!\Sq\}\uplus 
	p,\Rq\!::\!(f,\m,\many{v_r}))
	~\fut(f,\undefined) }

\inferrule[Invk-Passive]{
  \feval{e}{(\sigma+\ell)}=o \\ 
  \feval{\many{e}}{(\sigma+\ell)}=\many{v}\\
  \bind(o,m,\many{v})
  = \{\ell' \mid s'\}
}
{
  \act(\alpha,o_\alpha,\sigma,\{q\mapsto\{\ell \mid x=e.\m(\many{e});s\}::\Sq\}\uplus p,\Rq)
 \to \act(\alpha,o_\alpha,\sigma,\{q\mapsto\{\ell'\mid s'\}::\{\ell\mid x=\bullet;s\} 
 ::\Sq\}\uplus p,\Rq) 
}

\inferrule[Return-Local]{	v = \feval{e}{(\sigma+\ell)}
}
{
  \act(\alpha,o,\sigma,\{q\mapsto\{\ell \mid \key{return}~e;s_r\}::\{\ell'\mid 
  x\!=\!\bullet;s\}::\Sq\}\uplus p,\Rq)
  \to \act(\alpha,o,\sigma,\{q\mapsto\{\ell'\mid x\!=\!v;s\}::\Sq\}\uplus p,\Rq) 
}

\inferrule[Return]{
	v = \feval{e}{(\sigma+\ell)}  }
	{\act(\alpha,o,\sigma,\{(f,m,\many{v})\mapsto\{\ell \mid \key{return}\ 
	e;s_r\}\}\uplus p,\Rq) \ \fut(f,\undefined)\\\\
	 \to \act(\alpha,o,\sigma,p,\Rq) ~~ \fut(f,v,\serialise(v,\sigma))
	}
\qquad
\inferrule[Update]{
	\sigma(o)=f \\(v_r,\sigma_r)=\rename{\sigma}{v,\sigma'}\\\\
\sigma''=\sigma[o\mapsto v_r]\cup\sigma_r
}
{	\act(\alpha,o_\alpha,\sigma,p,\Rq)\
        \fut(f,v,\sigma')\\\\
      \to \act(\alpha,o_\alpha,\sigma'',p,\Rq) ~~ \fut(f,v,\sigma')
	}
		\end{footnotesize}
    \end{mathpar}
	\caption{Semantics of \multiasp.}
	\label{fig:translation-multiasp-semantics} 
\end{figure}

Figure~\ref{fig:translation-multiasp-semantics} shows the semantics of 
\multiasp as a transition 
relation between configurations. The rules only show the activities and futures involved 
in the 
reduction, the rest of the configuration is kept unchanged.  Most of the rules 
are triggered depending on the shape of the first statement of an activity's 
thread. The reduction rules are described below:

\begin{itemize}
	\item \textsc{Serve} picks the first ready request in the queue (i.e. compatible with 
	executing requests and with older requests in the queue) and allocates a new thread 
	to serve it.
It fetches the method body and creates the execution context.
	\item\textsc{Assign-Local} assigns a value to a local variable. If the statement to 
	be executed is an assignment of an expression that can be reduced to a value, then 
	the mapping of local variables is updated accordingly. 
	\item\textsc{Assign-Field} assigns a value to a field of the current object (the one pointed to by $\ethis$). It is similar to the previous rule except that it modifies the local store.
	\item\textsc{New-Object} creates a new local object in the store at a fresh location, after evaluation of the object parameters. The new object is assigned to a field or to a local variable by one of the two rules above.
	\item\textsc{New-Active} creates a new activity that contains a new active object. It 
	picks a fresh activity name, and assigns the serialised object parameters: the 
	initial local store of the activity is the piece of store referenced by the  
	parameters. The freshness of the location of the new active object ensures that it is 
	not  in the serialised  store.
	\item\textsc{Invk-Active} performs an asynchronous remote method invocation on an 
	active object. It creates a fresh future with undefined value. The arguments of the 
	invocation are serialised and put in the store of the invoked activity, possibly 
	renaming locations to avoid clashes. The special case $\alpha=\beta$ requires a 
	trivial adaptation with the rule \textsc{Invk-Active-Self}.
%
%

	\item\textsc{Invk-Passive} performs a local synchronous method invocation. The method 
	is retrieved and an execution context is created;  the thread stack is extended with 
	this execution context. The interrupted  execution context is second in the stack; 
	the result returned by the method  will replace the $\bullet$ when it is computed. 
	Note 
	that the distinction between synchronous and asynchronous calls depends on the kind 
	of the reference to
	the invoked object: an active object is invoked synchronously if it 
	is invoked through its local reference.
	\item\textsc{Return-Local} handles the return value of local method invocation. It 
	replaces the $\bullet$ in the second entry of the  stack by the returned value. The 
	return 
	 corresponds  to a local invocation because it is not the only execution 
	context in the stack.
	\item\textsc{Return} occurs when a request finishes. It stores the value 
	computed by the request as a future value. Serialisation  packs the objects 
	referenced by the future value.
	\item\textsc{Update} updates a future reference with a resolved value.  This is performed at any time when a future is referenced and the future value is resolved.
\end{itemize}

The local rules reflect a classical object oriented language: 
\textsc{New-Object} and \textsc{Assign-Field} modify the local store, and 
\textsc{Invk-Passive} and \textsc{Return-Local} affect the local execution 
context. The other rules deal with parallelism and communication.
Given a program 
$P= \vect{\C}\ \wrap{\vect{x}\ s}$, executing $P$ requires to create an initial 
configuration with a single activity that serves one request containing the main block 
$s$ with local variables $\vect x$: $\Cn_0=\act(\alpha,o,\emptyset,\{q\mapsto\{ 
\manyMap{x \mapsto \enull},\ethis\mapsto o \mid s\}\},\emptyset)$. Classically $\to^*$
	denotes the transitive closure of $\to$.
In~\cite{Henrio2013}, it was proven that this semantics ensures that request services are 
scheduled such that
\emph{parallelism is maximised while preventing two incompatible requests
	from being served in parallel}. Safe parallelism can be formalised as follows
\begin{theorem}[Safe parallelism]
	Any two requests served in parallel are compatible: if $\Cn_0$ is the initial 
	configuration and
	 $\Cn_0\to^* \act(\alpha,o,\sigma,\{q \!\mapsto\!\{\ell \!\mid\! s\}::\Sq\} 
	 \uplus \{q' \!\mapsto\!\{\ell' \!\mid\! s'\}::\Sq'\} \uplus p,\Rq) ~ Q$ with 
	 $q=(f,\m,\many{v})$, $q'=(f',\m',\many{v}')$ then 
	 $\Compatible(q,q')$.
\end{theorem}

\subsection{Request Scheduling Extension}
\label{sbsc:framework-scheduling}
The semantics presented above ensures that any ready request can be 
served immediately, which 
maximizes parallelism.
While this property is strong and valuable, such a scheduling policy might reveal 
inefficient in practice and some 
additional mechanism is needed to control the number of threads running in parallel. 
To control the parallelism the programmer can limit the number of 
executing threads globally or per group. We detail below the 
 annotations provided in \proactive for thread limitation and extend the semantics to 
 model this aspect.

\subsubsection{Controlling thread creation in ProActive}

To customise   scheduling inside multi-active objects we 
introduce  scheduling annotations. 
The \threadannot annotation allows the programmer to control two aspects:
\begin{itemize}
	\item \textbf{The maximum number of threads used by a multi-active object.} This is 
	the 
	\intro{thread limit} of the multi-active object. It prevents a potential thread 
	explosion at runtime. This limit is checked whenever a thread is to be created or 
	activated.
	\item \textbf{The kind of thread limit.} It can be of two kinds: either \intro{soft} 
	of \intro{hard}. A \intro{soft thread limit} counts, in the thread limit, only the 
	threads that are currently active, i.e. that are not blocked in wait-by-necessity. On 
	the contrary, a \intro{hard thread limit} counts, in the thread limit, all the 
	created threads, even the ones that are in wait-by-necessity.
\end{itemize}

\begin{lstlisting}[float, label=lst:scheduling-thread-config, caption=An example of 
annotations for thread  management.]
@DefineThreadConfig(threadPoolSize=2, hardLimit=false)
@Group(name="monitoring", selfCompatible=true, minThreads=1, maxThreads=2)
\end{lstlisting}

The \threadannot annotation is illustrated in Listing~\ref{lst:scheduling-thread-config}, 
Line 1. A multi-active 
object  created with this annotation  has two threads at 
maximum to process its requests. Since the 
configuration specifies a soft thread limit, additional threads can  be 
created to compensate threads that are blocked in wait-by-necessity. 
The thread limit and the kind of thread limit can also be changed programmatically and 
dynamically through the API. 

%

In addition to the global thread limitation that applies per multi-active 
object,  a mechanism  controls the number of threads  
allocated to a given group. 
This mechanism first  
 reserves some threads for the processing of the requests of the group. Second, it  
 specifies the maximum number of 
threads that can be used by the requests of the group at the same time. 
Since these specifications apply to groups of requests, we extend the \code{@Group} 
annotation with two optional parameters. 
Line 2 of Listing~\ref{lst:scheduling-thread-config}
 shows a group 
definition where one thread  is reserved for the 
 group and  this group cannot use more 
than two threads.

\begin{lstlisting}[float, label=lst:scheduling-priority, caption=An example of priority annotations.]
@DefinePriorities({
	@PriorityHierarchy({
		@PrioritySet({"gettersOnImmutable"}),
		@PrioritySet({"broadcasting", "dataManagement"}),
		@PrioritySet({"monitoring"})
	})
})
\end{lstlisting}

\medskip When the number of threads is limited, 
requests compete for thread resources. Due to the limited number of threads, 
there might exist requests ready for execution (because they fulfil 
compatibility conditions), but that cannot be executed because all threads of 
the multi-active object are busy. Such requests form the \intro{ready queue} of 
a multi-active object. By default, the ready queue is, like the reception queue, 
ordered according to the order of reception.  We introduce a priority mechanism 
 that allows the programmer to determine an order of 
importance in the execution of requests. Priorities applies on the ready 
request queue, and thus on compatible requests. We extend the set of 
multi-active object annotations with priority annotations that relate groups of 
requests. We represent the priority dependencies with a graph structure, that 
allows a partial ordering of requests. A \code{@PriorityHierarchy} annotation 
creates an ordering on groups. The order in which the groups are declared 
defines the priority dependencies, like \code{group1 > group2 > group3}. 
Several groups can belong to the same priority level. 
Listing~\ref{lst:scheduling-priority} shows an example of priority annotations applied to 
the \code{Peer} class of 
Listing~\ref{lst:application-peer}. It states that methods of the group 
\code{gettersOnImmutable} should be served in priority compared to other methods, 
\code{monitoring} methods have the lowest priority, and  there is no order
between the two other groups.

Overall, the advanced scheduling annotations for priorities and manipulation of 
threads offer a high-level specification mechanism to 
deal with fine-grained optimisation of multi-active object-based applications. The 
\proactive API also allows the programmer to switch between hard and soft thread limit 
at runtime, for the current active object.

\subsubsection{The semantics of request scheduling}
 We formalise in \multiasp the management of threads in \maos. Our 
approach is to introduce in the semantics additional qualifiers on activities 
and requests to record scheduling informations.  First, we 
extend the syntax of \multiasp so that the thread limit mechanism can be 
programmatically changed between a \emph{soft limit} and a \emph{hard limit}. The 
syntax is extended with two new 
statements:

\newcommand\setLimitSoft{\textbf{setLimitSoft}\xspace}
\newcommand\setLimitHard{\textbf{setLimitHard}\xspace}
\newcommand\sh{\text{\emph{{sh}}}}
\newcommand\Group{\nt{group}}
\mysyntax{\mathmode}{
\simpleentry s [] ... \bnfor \setLimitSoft \bnfor \setLimitHard  []
}

Then, we extend the 
\multiasp semantics with request scheduling aspects. We suppose
 that the group of a request $q$ can be retrieved 
through an auxiliary function $\Group(q)$. Additionally, a filtering operator 
$p\big|_{g}$ returns the requests from group $g$ among the set of threads $p$. 
The thread limit of a group $g$ can be retrieved with $\Limit{g}$. To
indicate the status of each thread, we qualify each of the currently served 
requests as either actively served: $q_A$, or passively served: $q_P$. Each entry in the 
current request queue is now 
either $q_A\mapsto\Sq$ or $q_P\mapsto\Sq$. Passively served requests  are 
requests that have been
blocked 
in wait-by-necessity. The auxiliary function 
$\NbActive{p}$ returns the number of actively served requests in the set of 
threads $p$.  
Each activity has either a soft limit status written 
$\act(\ldots)_S$, or a hard limit status written $\act(\ldots)_H$. An activity 
has by default a soft limit status when it is created. $\sh$ is used as a 
variable ranging over \small $S$ \normalsize and \small $H$\normalsize: $\sh::=S\,|\,H$.
Finally, we modify the reduction rules 
of the operational semantics as follows:

\begin{itemize}
	\item We add a rule \textsc{Activate-Thread} for activating a thread. It looks at the 
	group of the considered request and checks if this group has not reached its thread 
	limit. The $\sh$ variable is used so that the kind of thread limit is kept unchanged.
\mathsizebegin 
    \inferrule[Activate-Thread]
{\Group(q)=g\\ \NbActive{p\big|_g}<\Limit{g}}
{	\act(\alpha,o,\sigma,\{q_P \mapsto \Sq\}\uplus p,\Rq)_\sh
      \to \act(\alpha,o,\sigma,\{q_A \mapsto \Sq\}\uplus p,\Rq)_\sh }
\mathsizeend

	\item Each rule allowing a thread to progress requires that the request processed by this thread is active. To this end, $q$ is replaced by $q_A$ in all \multiasp reduction rules except for \textsc{Serve} and \textsc{Update}.
	\item The \textsc{Serve} rule is only triggered if the thread limit of the request 
	group is not reached, i.e. if: \small 
	$\NbActive{p\big|_{\Group(q)}}<\Limit{g}$\normalsize.
	\item We add two additional rules, \textsc{Set-Hard-Limit} and 
	\textsc{Set-Soft-Limit}, for changing the kind of thread limit of an activity:
\mathsizebegin
    \inferrule[\ \ \kern-0.5ptSet-Hard-Limit]{}
    {\ \ \act(\alpha,o,\sigma,\{q_A\!\mapsto\!\{\ell \mid \setLimitHard;s\}::\Sq\}\uplus 
    p,\Rq)_\sh
      \to \act(\alpha,o,\sigma,\{q_A\!\mapsto\!\{\ell\mid s\} ::\Sq\}\uplus p,\Rq)_H }
  
  \inferrule[Set-Soft-Limit]{}
  {\act(\alpha,o,\sigma,\{q_A\!\mapsto\!\{\ell \mid \setLimitSoft;s\}::\Sq\}\uplus 
  	p,\Rq)_\sh
  	\to \act(\alpha,o,\sigma,\{q_A\!\mapsto\!\{\ell\mid s\} ::\Sq\}\uplus p,\Rq)_S }
\mathsizeend


\item A  wait-by-necessity occurs only in 
  case of method invocation on a future, since field access is only allowed on the 
  current object $\ethis$. We add a rule \textsc{Invk-Future} that passivates the current 
  thread when a method invocation is performed on a future that has not been updated 
  locally yet. This rule is only 
  applied for activities in \emph{soft 
  thread limit} and when a method is invoked on a reference to a future.
\end{itemize}
\smallskip
\mathsizebegin
    \inferrule[Invk-Future]{ \feval{e}{(\sigma+\ell)}=o' \\ \sigma(o')=f} {
      \act(\alpha,o,\sigma,\{q_A\!\mapsto\!\{\ell \mid 
      x=e.\m(\many{e});s\}::\Sq\}\uplus p,\Rq)_S
      \to \act(\alpha,o,\sigma,\{q_P\!\mapsto\!\{\ell\mid x=e.\m(\many{e});s\} 
      ::\Sq\}\uplus 
      p,\Rq)_S }
\mathsizeend

\smallskip

In summary, four  rules are added to \multiasp semantics, and the terms for 
activities and requests are qualified with thread indicators  to take 
into account advanced scheduling mechanisms of \maos.
Note that the thread limit is only checked upon thread activation, thus when switching 
from soft to hard limit, there may be interrupted passive threads in the set of currently 
executed requests. In other words, it is not always true 
that, when the current limit is hard, there is no passivated currently executed request. 
This corresponds to the behaviour of \proactive that checks thread limit only upon 
creation 
or 
activation of a thread.
Though the semantics allows for activation/passivation loops for the same thread, in 
practice useless activations are avoided by monitoring the status of the 
futures and activating, by a notification mechanism,  a thread that can 
progress.

\subsection{Development Support}
\label{sbsc:framework-support}
We conclude this section by presenting a visualisation tool dedicated to multi-active 
objects. Its purpose is to help the programmer understand the behaviour of his 
multi-active object applications and to debug concurrent programs  easily.

\subsubsection{Viewing executions}
The multi-active object programming model is designed such that there is a 
minimum specification to be written by the programmer. However, multi-active 
objects are  very good for programming systems that involve complex 
coordination of entities and massive parallelism. For these advanced cases,
being able to observe the behaviour of the application is crucial, either to debug it or 
to improve its performance.

We present a debugger that offers a visualisation of multi-active object 
executions based on a post mortem analysis. The user of this tool can observe the
application execution while being exposed to a view that corresponds to the  notions he 
manipulates when programming. In the main frame, a thread is 
represented by a 
sequence of requests on a time line. A screenshot of the debugger tool is 
provided in Figure~\ref{fig:support-deadlock-1}. In this example, two single-threaded
multi-active objects are displayed. Arrows represent the 
request sending between multi-active objects. The length of an element is 
representative of its duration. A mouse-over on a request displays the name of 
the request, the identity of the sender, and the timestamp of its reception.
 The debugger highlights the 
compatibility of the selected request with respect to others. A precise listing of 
requests can also be viewed based on a particular 
timestamp, distinguishing the requests that are being executed, the ones in the 
queue, the completed ones, and the ones that will be received later.


\subsubsection{Use case}

\begin{figure}[t]
	\center
	\includegraphics[width=.97\textwidth]{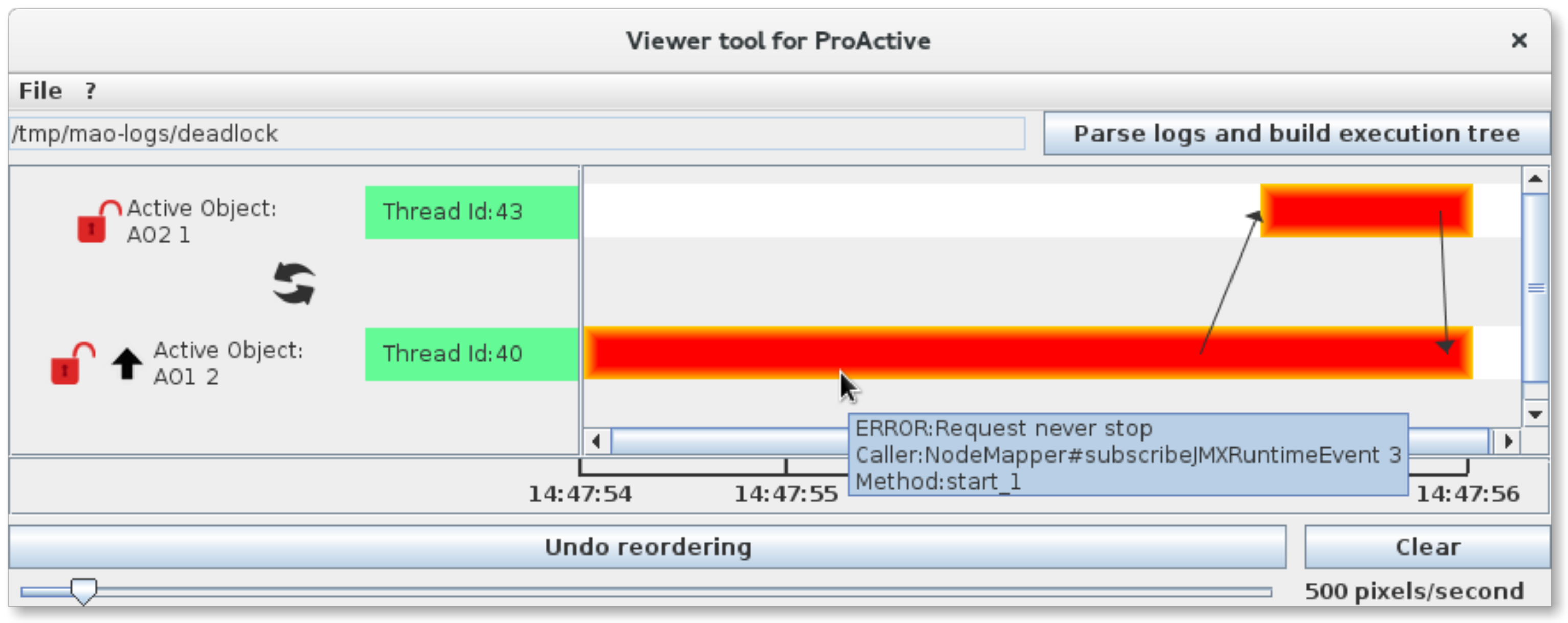}
	\caption{The debugger tool: deadlock scenario (screenshot).}
	\label{fig:support-deadlock-1}
\end{figure}

In active objects, a deadlock is generally caused by a circular dependency of 
requests. With multi-active objects, a deadlock can be due to two 
constraints: a lack of compatibility or a lack of threads. We show in 
Figure~\ref{fig:support-deadlock-1} the debugger tool in a scenario where it helps the 
programmer to identify a deadlock in a multi-active 
object execution (red requests). The debugger produces in this case the 
`request never ends' error, informing the user of a potentially faulty execution. The 
representation of communications shows that there is a circular request 
dependency between multi-active objects \code{First} and \code{Second}. 
For example, if the request received by \code{First} was not 
compatible with the request that executes, then it would never execute. 
Here the deadlock is instead due to a lack of threads:  no more thread can be 
created. This case only happens if the soft 
thread limit is not activated,  else another thread could 
compensate the thread that is in wait-by-necessity.

Besides deadlocks, the debugger also shows the sequence of 
actions that occurred in a particular execution, and thus help in spotting 
race conditions that can happen, typically the concurrent sending 
of requests to the same target object that would create non-deterministic behaviour. This 
tool is part of the \proactive framework.

\section{Encoding of Cooperative Active Objects into Multi-active Objects}
\label{sec:encoding}

We present in this section a particular application of multi-active objects that 
consists in encoding the cooperative active object language \absl into the 
multi-active objects of \multiasp/\proactive. In other words, we present here a 
\proactive~backend for \absl that translates \absl code to Java programs that 
use the \proactive library and offers a distributed execution of \absl programs. We also 
formalise
and prove the correctness of the translation from \absl to \multiasp programs. 

\subsection{\proactive Backend and Translational Semantics}
\label{sbsc:encoding-backend}
One way to execute \absl programs is to use the Java backend. The Java backend
translates ABS programs into Java programs according to \absl semantics, but without 
distribution support. The \proactive backend 
relies on  the Java backend concerning the functional layer of 
the language, introducing a new translation for the object and 
concurrency layer, plus distributed notions.

First, to translate the object model of \absl into multi-active 
objects, we put several objects under the control of one active object, which 
fits the active object group model of \absl. Compared to translating each 
\absl object into a multi-active object, this solution 
requires less synchronisation between activities and is more scalable. Indeed to 
guarantee that a single 
thread executes in a set of active objects we would need to synchronise the different 
active objects and schedule their execution. Implementing such a scheduling pattern would 
not only be difficult and costly (each synchronisation should be performed either by a 
method invocation or by a future access) but also would not match the active object 
principles (loose coupling between active objects). 
In the translation, we introduce a 
\COGClass class for representing \absl \COGS. Only objects of the \COGClass 
class are multi-active objects. The \COGClass class in \multiasp is shown in 
Listing~\ref{lst:encoding-cogclass}. \texttt{UUID} represent the type of object 
identifiers. The \COGClass class has methods to store and 
retrieve locally the translated \absl objects, and to generically execute a method 
on them. A special method \code{execute\_condition} is used to execute await 
conditions. It is a separate method because it does not have the same purpose as 
\code{execute} that serves external asynchronous requests; and the two methods 
 follow different compatibility rules.
We assign to every 
translated \absl object a unique 
identifier. 

\begin{lstlisting}[float, label=lst:encoding-cogclass, caption=The  COG class in MultiASP, mathescape=true]
Class COG {
  UUID freshID() 
  UUID register(Object x, UUID id)  
  Object retrieve(UUID id) 
  Object execute(UUID id, MethodName m, $\many{\nt{params}}$) {
  	w=this.retrieve(id); x=w.m($\many{\nt{params}}$); return x}
  Object execute_condition(UUID id, MethodName cm, $\many{\nt{params}}$) {
   	w=this.retrieve(id); x=w.cm($\many{\nt{params}}$); return x}
}
\end{lstlisting}  
  
Second, all translated \absl classes are extended with two parameters: a   
$\nt{cog}$ parameter, storing the \COG to which the object belongs, and an 
$\nt{id}$ parameter, storing the identifier of the object in that \COG. The 
methods $\nt{cog}()$ and $\nt{myId}()$ return those two   parameters. A dummy 
method $\nt{get}()$ that returns \code{null} is also added to each object, it is used to 
perform synchronisation on future objects.





Concerning statements, the translation only 
impacts the statements that  deal with object 
creation, method   invocation, or future manipulation. The 
translation of \absl statements and expressions into 
\multiasp is shown in Figure~\ref{fig:Transl}, and explained below. The code 
that is generated by the \proactive backend for \absl is similar to what is 
shown in \multiasp. We organise the description as follows. First we describe the storage 
and access to objects, this explains the translation of the two \enew statements. Then, 
 we define the translation of method invocations 
and particularly of asynchronous method calls. Section~\ref{sec:coop-sched} defines 
synchronisation on futures and  cooperative scheduling aspects. Requests are split 
into three groups: one for executing requests, one for 
allocating fresh object identifiers, and a third one for 
registering objects in a \cog. Groups and compatibility are presented in 
Section~\ref{sec:groupsABS}. 
 Section~\ref{sec:await-expr} describes \await\ statement on boolean expressions.

\addtolength{\textfloatsep}{-.5ex}
\begin{figure}[!tp]
	\centering 
	\renewcommand{\arraystretch}{0.9} 
	\vspace{-2ex}

	\begin{footnotesize}
		\begin{math}
\begin{array}{@{}ll@{}}
\begin{array}{@{}l}
\llbracket x\!=\!\enew {\C(\many e)}\rrbracket\triangleq\begin{array}[t]{@{}l}
\nt {newcog}\!=\!\newActive{\COGClass()};\\
\nt {id}=\nt {newcog}.\nt{freshId}();\\
\nt no=\enew C(\many e,\nt {newcog},\nt{id});\\
z=\nt {newcog}.\nt {register} (\nt{no},\nt{id});\\
x=\nt{no}
\end{array}
\\\\[-1ex]
\llbracket x\!=\!\enew {{local}\ \C(\many 
	e)}\rrbracket\triangleq\!\begin{array}[t]{@{}l}
t = \ethis.\nt{cog}();\\
\nt {id}=t.\nt {freshId}();\\
\nt {no}=\enew \C(\many e,t,\nt{id});\\
z=t.\nt {register} (\nt{no},\nt{id});\\
x=\nt{no}
\end{array}
\\\\[-1ex]
\llbracket x=e!\m(\many e)\rrbracket~\triangleq~
\begin{array}[t]{@{}l@{}}
t=e.\nt{cog}();\nt {id} =e.\nt{myId}();\\
x=
t.\nt{execute}(\nt{id},\m,\many e)
\end{array}	
\\\\[-1ex]
\llbracket x\!=\!e.\m(\many 
e)\rrbracket\triangleq\begin{array}[t]{@{}l@{}}
t=e.\nt{cog}();b=\ethis.\nt{cog}();\\
\nif (t==b)\quad \{x=e.\m(\many e){}\}\\
\nelse~ \{\!\begin{array}[t]{l}

\nt {id}=e.\nt{myId}();\\
x=t.\nt{execute}(\nt {id},\m,\many
e);\\\setLimitHard;\\
w=x.\nt{get}();\\
\setLimitSoft  \}
\end{array}
\end{array}


\end{array}	
&
\begin{array}{l}
\llbracket\nt x =\nt e\rrbracket~\triangleq~x =\nt e
\\\\[-1ex]
\llbracket\nt x =\nt 
y.\nt{get}\rrbracket\triangleq\begin{array}[t]{@{}l@{}}
\setLimitHard; \\
\nt w=y.\nt{get}();\\
\setLimitSoft;\\
\nt x=\nt y
\end{array}
\\\\[-1ex]
{ \llbracket await~ g\rrbracket\triangleq\!	\begin{array}[t]{@{}l@{}} \nif 
(\lnot 
g) \quad \{~\\
	\quad
	\begin{array}[t]{l}
	\nt t=\ethis.\nt{cog}();\\ \nt {id}=\ethis.\nt{myId}();\\
	z=\nt t.\nt{execute\_condition}(\nt{id},\nt{condition}\_g,\many{x});\\
	w=z.\nt{get}() ~\}
	\end{array}\end{array}	}\\
\quad\text{where $\many x$ are the local variables used in $g$}\\
\quad\text{and $g$ 
contains 
no future 
await of the form $y?$}
\\\\[-1ex]
\llbracket suspend\rrbracket \triangleq \begin{array}[t]{@{}l@{}}
\nt t=\ethis.\nt{cog}();\\ \nt {id}=\ethis.\nt{myId}();\\
z=\nt t.\nt{execute\_condition}(\nt{id},\nt{condition}\_\nt{True},\many{x});\\
w=z.\nt{get}()
\end{array}
\\\\[-1ex]
\llbracket\nt{await}\ 
x?\rrbracket~\triangleq~w=x.\nt{get}()
\\\\[-1ex]
\llbracket\nt{await}\ g\land g'\rrbracket\triangleq\llbracket\nt{await}\ 
g\rrbracket;\llbracket\nt{await}\ g'\rrbracket
	\end{array}		
	\end{array}	
		\end{math}
		\end{footnotesize}
\vspace{-1.2ex}
	\caption{\label{fig:Transl} Translational semantics from ABS to
		\multiasp} 
\end{figure}

\subsubsection{Object addressing}
To allow all objects to be accessible like in \absl, we use a 
two-level reference system. Each 
\COG is accessible by a global reference and  each \absl 
object is accessible inside its \COG through a local identifier. The pair 
(\COG, identifier) is a unique reference for all translated \absl objects and 
allows the runtime to retrieve any object.

We translate the \absl \code{new} statement ($\llbracket x\!=\!\enew {\C(\many e)} 
\rrbracket$), that 
creates a new object in a new \COG, with the instantiation of a 
\COG multi-active object and of the new object in \multiasp. First, the \COG 
multi-active object is created with the \code{newActive} primitive. Thus, 
further method invocations on the $newcog$ variable will be asynchronous remote 
method calls. Then, a fresh identifier is retrieved from the new \COG, 
to create the new object that is stored in a reserved temporary local variable 
$\nt no$. The new object is a passive object that has a reference to its \COG. 
Then, the new object is referenced in the new \COG through the \code{register} 
method. Since the new \COG is a remote multi-active object, the new object is 
copied when it is transmitted to the new \COG through the \code{register} 
invocation. Finally, the \code{x} variable references the new object that is 
ready to be used.

For objects that are created with \code{new local} in \absl ($\llbracket x\!=\!\enew 
{{local}\ C(\many e)} \rrbracket$), the translation in \multiasp is similar to the 
translation of the \absl \code{new} statement, but the local \COG is retrieved 
instead of creating a new one. No new activity is created here and the object is 
registered locally, without copy.

\subsubsection{Method invocation}
  
 When an 
asynchronous method call is performed in \absl ($\!e!\m(\many e)$), in 
 \multiasp a remote method 
invocation is executed. Because of the hierarchical object referencing, we first retrieve 
the \COG of the   
translated object that is invoked in \absl, as well as the identifier of the 
object. Then, we perform a generic method call (implicitly asynchronous) named 
\code{execute}, on the retrieved \COG. The \code{execute} method of the 
\code{COG} class looks up the targeted object through its identifier and runs on 
it the given method by reflection, with the translated parameters. 
In  \multiasp, an asynchronous method call entails a copy of the 
invocation parameters. Consequently, several copies of the same 
translated 
\absl object exist in \multiasp, whereas only one copy 
of each object exists in \absl. This is not a problem because only one of these 
copies is registered and  when an 
object's copy is used in the translated program (i.e. a method  is invoked 
on it\footnote{Field access outside the current object is not allowed in 
\absl\ and ASP.}), the same process is always applied: the invocation is forwarded to 
the \COG 
that manages the only registered object. 
Thus, in the end, only this registered object is manipulated by all the 
invocations made on the object's copies. Consequently, the behaviour by 
reference of \absl (and other similar active object languages) can be simulated 
with the behaviour by copy of  \multiasp. The same mechanism is 
 applied to handle the translation of future updates.
Also,  since the copies of an object are only used  to retrieve 
the registered object, all transmitted objects between \COG can be lightweight: 
they only need to embed the reference to the \COGClass and the identifier of the 
only registered object. In the  \proactive 
backend, we tune the serialisation mechanism so that only those fields are 
transmitted between activities,   saving memory and bandwidth in the execution of the
translated program .

For the case of synchronous method calls, in \absl ($!e.\m(\many e)$), 
we distinguish two cases, like in the \absl semantics. Either the  	
call is local and an execution context is pushed in the stack, or the call is 
remote and we perform an asynchronous  method   
invocation and immediately wait the associated future, blocking the current activity 
thread (see below).
  
\subsubsection{Cooperative scheduling} \label{sec:coop-sched}
%

First, we consider the translation of \absl \code{await} statements  on future variables 
($\llbracket \nt{await}\ x? \rrbracket$).
In \absl, an await statement on an unresolved future makes the current request release the execution thread.
In order to stop the execution of the request in the multi-active object-based 
translation, we trigger a wait-by-necessity.
To this end, we call the dummy \emph{get()} method so that a wait-by-necessity is 
automatically triggered if the future is unresolved. 
However, in  \multiasp, a wait-by-necessity blocks the thread and does not release it.
To encode cooperative scheduling we configure the \COGClass class with the following 
annotations configuring its internal scheduling.
Since in the translation, all remote method invocations go through the \code{execute} 
method of the \COGClass class, we put this method in a multi-active object group.
We declare this group self compatible so that several \code{execute} methods can be 
scheduled on several threads of the multi-active object at the same time.
Then, we set a thread limit of one thread and we use a soft thread limit:
\lstset{numbers=none}
\begin{lstlisting}
@Group(name="scheduling", selfCompatible=true)
@DefineThreadConfig(threadPoolSize=1, hardLimit=false)
\end{lstlisting}
This configuration first allows a single execute method to run at any time, so that there 
cannot be more than one running thread in a \COG multi-active object. 
The soft thread limit also allows a thread to process an \code{execute} request 
while  another \code{execute} request is passivated, waiting for a future.   Finally, as 
the translation of the await statement triggers a 
wait-by-necessity,
this setting ensures exactly the synchronisation of multi-active object threads that 
simulates the cooperative scheduling of \absl through the \code{await} statement.

Second, we consider the translation of \absl \code{get} statements ($\llbracket\nt x =\nt y.\nt{get}\rrbracket$).
In \absl, a \code{get} statement  retrieves a 
future value, blocking the execution thread if necessary. Compared to \code{await}, 
another thread cannot continue executing while 
waiting for a future through a \code{get} statement. In the translation, we  
temporarily go from a soft thread limit to a hard 
thread limit so that no other thread can start while the current thread is  waiting for 
the future to be resolved. To this end, we use the 
\setLimitSoft and \setLimitHard statements introduced in 
Section~\ref{sbsc:framework-scheduling}. In the translation, 
a \setLimitHard statement precedes the call to 
the \emph{get()} method, so 
that no other thread starts or resumes while performing the \code{get} operation. Then,  
the soft thread 
limit is restored.

\subsubsection{Groups and compatibility}  \label{sec:groupsABS}
In general,  to simulate the scheduling of \absl executions,  we create different method 
groups in the \COGClass class, and each group has its own thread 
limit:  
\begin{small}
\[
\begin{array}{@{}c@{}}
\Group(\nt{freshId})=g_1 \qquad \Group(\nt{execute})=g_2 \qquad \Group(\nt{register})=g_3 
\\ 
\Limit{g_1}=1 \qquad \Limit{g_2}=1 \qquad \Limit{g_3}=\infty 
    \end{array}\]
\end{small}
The compatibility relationship is defined such that\footnote{We use $\_$ as a wildcard: 
$q=(\_,\m,\_)$ means that there is 
	a future and a set of parameters such that $q$ has this form, in other words ``$q$ is 
	a 
	request on method \m.}:
\begin{small}\[
\Compatible(q,q')\iff\Big(\,\begin{array}{@{}l@{}}
\left(q = (\_, \nt{freshId},\_)~\Rightarrow~
q' \neq (\_, \nt{freshId},\_)\right) ~\land 
\\(\nexists id,x,\m,\many{e}.~
  q=(\_,\nt{register},[x, \nt{id}]) \land 
  q'=(\_,\nt{execute},[\nt{id}, \m, \many{e}])) 
    \end{array}\,\Big)\]
\end{small}
Group $g_1$ encapsulates \nt{freshId} requests. These requests cannot execute in parallel 
safely, so $g_1$ is not self compatible. Group 
$g_2$ gathers \nt{execute} requests (i.e. \absl requests). It is limited to one thread to 
comply with the threading model of \absl, and the requests are self compatible to enable 
interleaving.  Group $g_3$ contains \nt{register} requests that are self compatible and 
that 
have no thread limit. Concerning compatibility between groups, $g_1$ is  
compatible with other groups. Concerning $g_3$ and $g_2$, their compatibility is defined 
dynamically such that 
an \nt{execute} request and a
$register$ request are compatible if they do not affect the same identifier.
  \code{execute\_condition} methods are compatible with everything.
  
\subsubsection{\await\ on conditional expressions}  \label{sec:await-expr}
Regarding the translation of the \code{await} \emph{on conditions} and of the 
\code{suspend} statement of \absl, we realise it by the means of an additional 
multi-active 
object group and of a special configuration of its thread limit.
 First, note that  futures are single-valued assignments, and once they are available 
 they will remain available. Second, as defined in~\cite{Hahnle2013}, the \absl 
 \code{await} statement accepts only monotonic guards and conjunctive composition.
 Because of monotonicity, the translation of conjunctive \absl conditional guards 
 ($\llbracket\nt{await}\ g\land g'\rrbracket$) can be performed sequentially. Concerning 
 futures, this can be done by calls to the 
 \emph{get()} method of the translated \absl objects (the activity is in soft
limit at this point).
Then, in order to translate \absl conditional guards ($\llbracket \nt{await}~ 
g\rrbracket$), for each guard $g$, we generate a method \nt{condition\_g} that 
takes as parameters the variables $\many{x}$ used in $g$. A 
condition evaluation $g$ is defined as follows:
\[
   \begin{array}{@{}c@{}}
  \nt{condition\_g}(\many{x}) = \nwhile(\lnot g) ~ skip ; \ereturn\ \enull 
  \end{array}\]

In more recent semantics of \absl, guard are not monotonic; to encode this semantics, we
would have to check again the boolean part of the condition when the thread is resumed
(at this point no other thread can interfere with the object state). Each conjunctive
guard should be re-ordered so that it starts by a set of guards on futures (monotonic by
nature), and finishes with a  boolean expression guard. This boolean part of the await 
would be the only non-monotonic part that
should be checked again when the service thread is resumed.

We translate the \absl \code{suspend} statement ($\llbracket \nt{suspend}\rrbracket$) 
similarly to conditional guards, but with a condition that is always true. We define an 
\nt{execute\_condition} method in the \COGClass class that generically executes generated 
condition methods. The \nt{execute\_condition} method has its own group with an infinite 
thread limit because any number of conditions can evaluate in parallel (those requests 
are compatible with all 
the other 
requests):
\[
  \begin{array}{@{}c@{}}
    \Group(\nt{execute\_condition})=g_4
	\quad
    \Limit{g_4}=\infty
   \end{array}\]

\subsubsection{Distribution}

In addition to the translation of active object paradigms, the \absl compiler is slightly 
modified in order to enable distributed execution of \absl programs. We also make some 
other adaptations in the generated Java classes  to ensure an efficient distributed 
execution.

	\noindent\textbf{Serialisation.} The main challenge when moving objects from one 
	memory 
	space to 
	another is to reshape them so that they can be transmitted on the network medium. A 
	serialised version of an object must be created by the sender so that 
	 the object graph can be rebuilt from the serialised version by the 
	recipient. As \proactive is based on Java RMI, all objects that are part of a remote 
	method call (i.e. parameters and return values) must implement the Java 
	\code{Serializable} interface, otherwise a distributed execution throws an exception. 
	Thus,  the generated Java classes implement this 
	interface. 
	
	\noindent\textbf{Copy Optimisation.} To minimise copy overhead, in the 
	\proactive 
	translation we declare the 
	translated class fields with the \code{transient} Java keyword, which prevents them 
	from being embedded in a serialised version of an object (they are replaced by 
	\code{null}). The  fields that receive a value when the object is created go 
	through a customised serialisation mechanism: they are copied only the first 
	time they are serialised (i.e. when the object is copied into its hosting \COG). 
	After this initial copy, we only  copy the object
	identifier and the 
	reference to its \COG.
	We could also represent all the objects as generalised 
	references and store the object's state in a dedicated place. However the 
	translation of local synchronous calls to the original object would become 
	highly inefficient (compared to the current Java method invocation). This would also 
	make the 
	formalisation more complex.
	
	\noindent\textbf{Deployment.} Any distributed program needs a deployment 
	specification 
	mechanism 
	in order to place pieces of the program on different machines. \proactive embeds a 
	deployment descriptor that is based on XML configuration files, where the programmer 
	declares physical machines to be mapped to virtual nodes. Such virtual nodes can then 
	be used in the program in order to deploy an active object on them. In our 
	case, as the \proactive program is generated, we have to raise the node specification 
	mechanism at the level of the \absl program. We slightly modified the \absl syntax 
	(and parser) to allow the programmer to specify the name of the node on which a new 
	\COG must be 
	deployed. The \proactive backend then links this node name 
	to the deployment descriptor of \proactive. Now, an \absl \code{new} statement is 
	optionally followed by a string that identifies a node, as follows: 
\begin{lstlisting}
Server server = new "mynode" Server();
\end{lstlisting}
During translation, the \proactive node object corresponding to this node identifier is 
retrieved, and given as parameter of the \code{newActive} primitive to deploy the active 
object on this node. For this specification to work, a simple descriptor file  similar to 
the following one must be created and attached to the \absl program:
\begin{lstlisting}
<GCMDeployment>
  <hosts id="mynode" hostCapacity="1"/>
  <sshGroup hostList="172.16.254.1"/>
</GCMDeployment>
\end{lstlisting}
In this example, the virtual node  \code{"mynode"} is mapped to the machine with the IP 
address mentioned in the \code{hostList} attribute. DNS names can be specified here as 
well. Many other deployment options can also be defined~\cite{Baduel2006}. If several 
machines are 
specified in the \code{hostList}, then the \proactive backend picks one machine in the 
list each time a new \COGClass is created, in a round robin manner. 
If no node  is specified in \absl, then the new \COG is created on the same 
machine in a different JVM, enforcing  a strict isolation of \COGS.

\subsubsection{Wrap up and applicability}

In conclusion, by carefully setting multi-active object annotations, and by adapting to 
distribution requirements, we are able to execute \absl programs in a distributed setting 
by using \proactive. This translation is automatically handled by the 
\proactive backend for \absl. The approach presented here and instantiated in the case of 
\absl and \multiasp could be generalized and applied to other active object
languages, and systematically provide a way to deploy and run most active object 
languages. The
effort to port our result to other active object languages depends on the target 
language. 
To get an efficient translation of different active object models into distributed 
multi-active objects, one needs to answer the questions related to the object model and 
to the scheduling model of the active object language. Most of all, one must carefully 
consider the location of objects: ``Should objects be grouped to preserve the performance 
of the application? If yes, how and under which control?''. 
When it comes to distributing active objects over several memory spaces, the only
scalable solution that we have seen is to address the objects hierarchically. 
This strategy is 
easily applicable to any active object language based on object groups.
For example, adapting this work to \jcobox should raise no technical difficulty. 
The most challenging aspect is that in \jcobox the objects share
a globally accessible and immutable memory. In this case, the global memory could be translated
into an active object that holds all immutable objects: since they are immutable, 
communicating them by copy is correct.
In the case of uniform active object languages, like \creol, creating one active object per
translated object handles straightforwardly the translation but limits scalability. The best approach is to group several objects behind a same active object for performance
reasons, like building abstract object groups that resemble \absl and \jcobox. In the 
case of \encore, objects are already separated into active and passive objects, which 
makes the translation of the object model easier. Once the distributed 
organisation of objects has been defined, then preserving the semantics of the source 
language relies on a precise
interleaving of local threads, which is possible thanks to the various threading controls 
offered by multi-active objects. 
Simulating policies different from cooperative scheduling is also possible with 
multi-active objects.
For example, the transposition of the \proactive backend to \ambient could seem tricky on the scheduling aspect, due to the existence of callbacks. However, a
callback on a future can still be considered as a request that is immediately executed in parallel,
but  starts by a wait-by-necessity on the adequate future. 

\subsection{Experimental Evaluation}
\label{sbsc:encoding-evaluation}

We evaluate the \proactive backend for \absl on several \absl applications and we 
compare the   execution of the 
\proactive program generated by the \proactive backend to the execution of the 
program generated by the Java backend for \absl. 
We  first test four example programs given in the ABS tool 
suite. These  examples involve a bank account program that consists of 167 
lines of \absl and that creates 3 \COGS, a leader election algorithm over a 
ring (62 lines, 4 \COGS), a chat application (324 lines, 5 \COGS), and a 
deadlock example that hangs by circular dependencies between activities (69 
lines, 2 \COGS). For all examples, we observe that the behaviour of the program 
translated with the \proactive backend is the same as the one translated  with 
the Java backend.  Thus, the scheduling policy enforced in 
\proactive faithfully respects the one of the reference implementation. These 
examples  run in a few milliseconds, thus they are inadequate for performance 
analysis of distributed executions.

We focus the rest of the experiments on an application that requires 
more computational resources: the pattern matching of a DNA sequence in a 
database. We implement a 
MapReduce programming approach~\cite{Dean:2008:MSD:1327452.1327492} with one \COG per 
worker.  We  search a pattern of 250 bytes in a database of 5MB of DNA 
sequences. We compute the global execution time 
when varying the number of workers. When executing the program given by the Java backend, 
we execute 
it on a single machine; when using the 
\proactive backend, we deploy two multi-active objects (i.e. two workers) per 
machine. We use a cluster of the Grid5000  
platform~\cite{Cappello:2005:1542730}, where machines have 2 CPUs of 2.6GHz 
with 2 cores, and 8GB of memory.

\begin{figure}[t]
		\begin{subfigure}[b]{0.49\linewidth}
                \includegraphics[scale=0.37]{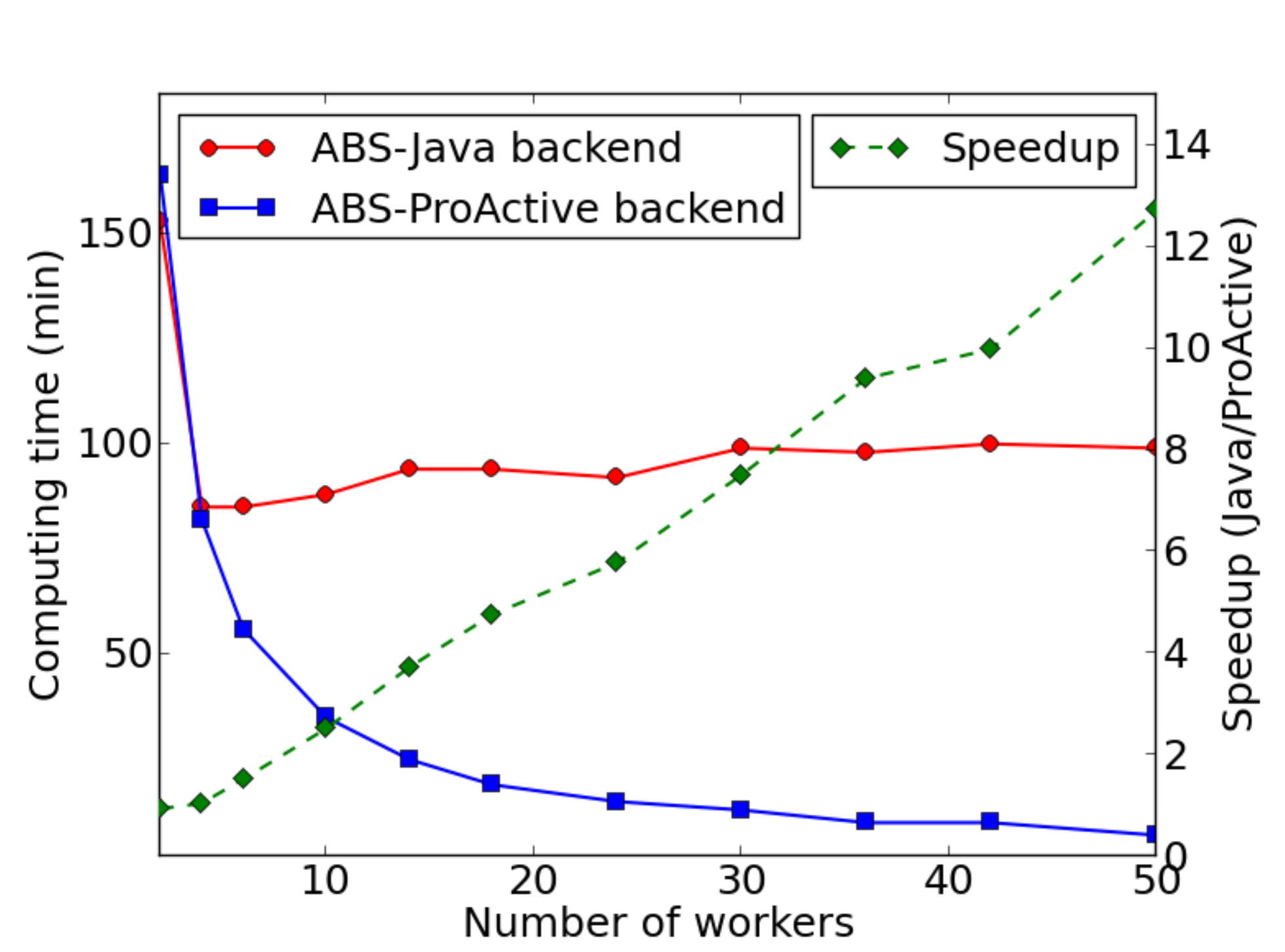}
                \caption{Against existing solution (time in minutes)}
                \label{fig:bench-usecase}
        \end{subfigure}
        \begin{subfigure}[b]{0.50\linewidth}
                  \includegraphics[scale=0.37]{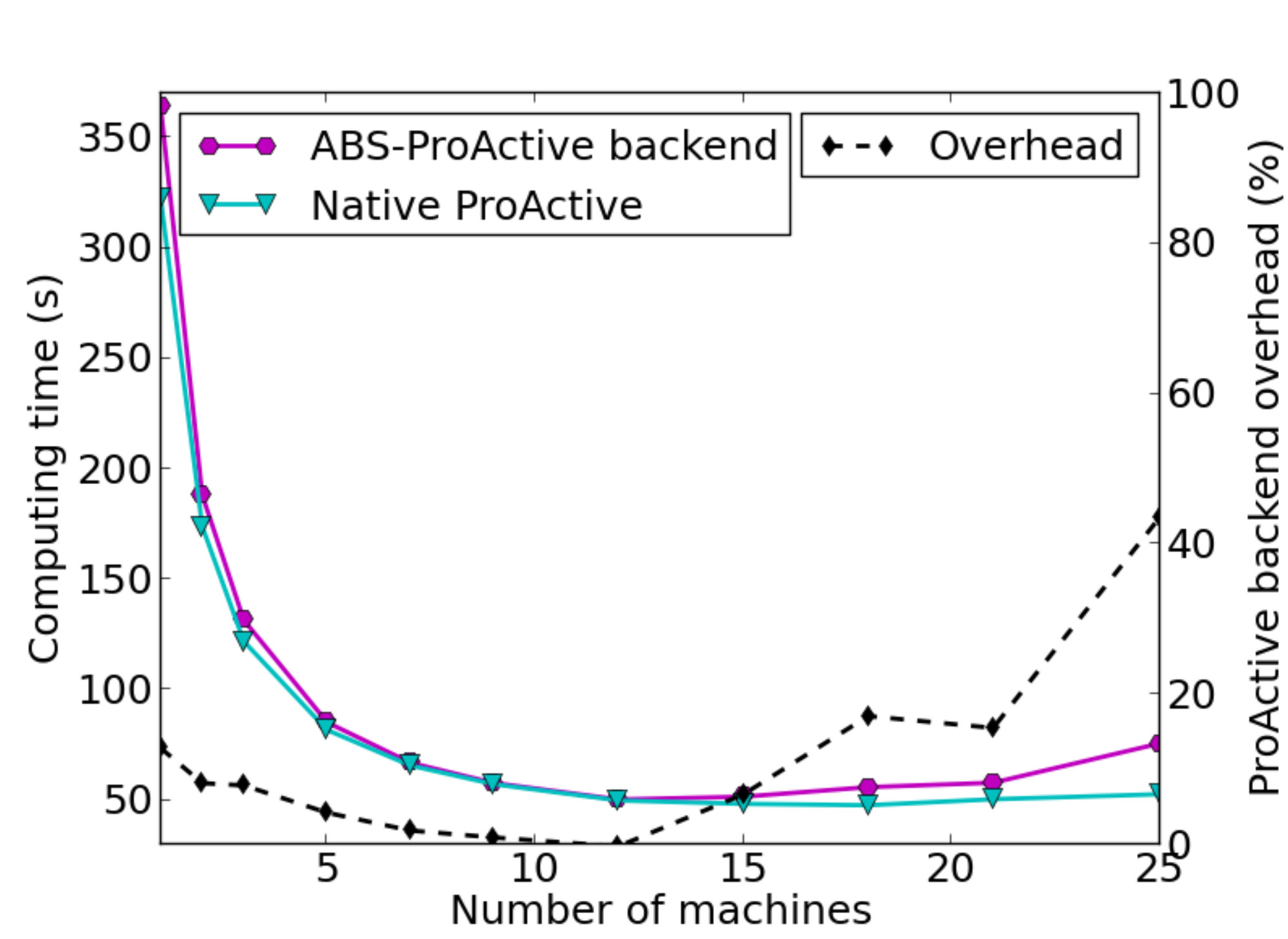}
                \caption{Against native code (time in seconds)}
                \label{fig:bench-native}
        \end{subfigure}
 \caption{Execution time of DNA-matching ABS application} 
\label{fig:graph}
\end{figure}

Figure~\ref{fig:bench-usecase} shows the execution time of both \proactive and 
Java translations of the \absl application from 2 to 50 workers.  The \proactive backend 
scales as expected. The 
execution of the Java program reaches an optimal degree of parallelism for 4 
workers (the number of cores of the machine) and then cannot 
benefit from higher parallelism. With the \proactive backend and 25 machines, the 
application completes in 19 minutes. 

We further evaluate whether the \proactive backend is effective compared to a 
program that would be directly written in \proactive.  Figure~\ref{fig:bench-native} 
compares the execution time of the two implementations of the DNA matching 
application: the first one is the generated program of the \proactive backend 
for \absl, and the second one is a hand-written version of the application in 
\proactive, i.e. executing according to \proactive semantics. 
Here, the program generated by the \proactive backend has been slightly modified: we have 
manually replaced the 
translation of functional \absl types  (integers, booleans, lists, and maps) 
with the corresponding standard Java types. Indeed, the implementation of the 
functional layer of \absl in Java is not as efficient as primitive Java types. 
Considering that our point is to evaluate the additional communication cost of 
the \proactive translation, we do not take into account the implementation of 
\absl types in Java. The performance is 
approximately 30 times better depending on the implementation of types. 
This experiment can be split into two parts: when the number of machines grows up to $12$ 
some speed-up can be noticed in both implementations, while above $12$ 
machines the time spent in communication becomes too important compared to the 
computation time, and no  significant performance improvement is obtained when adding 
more machines.
While the 
amount 
of computation performed by each peer is significant,  
the overhead introduced by the \proactive backend is rather low compared to a 
native version of the application, since it is overall under 10\%. At 
a larger scale, the generated version is characterised by additional 
communications to access intermediate objects and thus  the translated 
application suffers even more from the time wasted in communication; it becomes less 
efficient. 

To conclude, with the 
\proactive backend, one can run efficiently high-performance distributed application 
written in \absl and 
benefit from 
distribution compared to the Java backend. We conducted  experiments 
with a 
new backend for Java 8 on the same applications and reached a similar conclusion. 
This comparison is based on a ``high-performance computing'' application. An application 
relying on more 
thread interleaving, and coordination of interleaving threads 
would probably be less efficient in the \proactive backend. 
 

\subsection{Formal Study of the Translation}
\label{sbsc:encoding-formalisation}
This section investigates the  equivalence between \absl programs and 
their translation in \multiasp. 
We present two theorems stating under which conditions the \absl program and its 
\multiasp translation are equivalent. More precisely, we have one theorem for each 
direction of the simulation. We achieve the proofs of the theorems in terms of weak 
simulation in the two cases.
We formally prove that, except from some reasonable 
differences between the two languages, the translated program behaves exactly like the original one. The differences 
are mostly due to the distributed nature of the execution, and to the fact that futures 
have a different semantics in the two languages. It would have been possible to overcome 
those differences by providing an encoding that is more artificial and less shallow 
(adding intermediate objects or threads) but this would have prevented us from drawing 
conclusions about the  similarities and differences between the languages. 
Figure~\ref{fig:abs-runtime} recalls the runtime syntax of \absl. It defines the 
structure of \absl runtime configurations. Compared to \multiasp\ configurations, 
the structure of objects and futures are relatively similar except that each object has 
an entry in the configuration and thus active objects and futures do not have a local 
store; 
additionally request 
queues consist of already instantiated method bodies.  A 
\cog\ element is used to denote which object of each \COG\ is the active one.

The semantics of \absl is shown in 
Figure~\ref{fig:abs-semantics}, it uses a function \bind\ and an evaluation function 
$\feval{}{}$ similar to the ones of \multiasp. 
It uses two reserved variables in the local stores: \emph{destiny} stores the identifier 
of the future for which the current computation is performed; \emph{cont} stores the 
identifier of the future of the request that was interrupted in order to perform a local 
synchronous method invocation; this request should be recovered upon reaching the 
\key{return} statement. The rest of the semantics is quite similar 
to~\cite{Giachino2015}. The asynchronous 
invocation enqueues a new request and creates a new futures. Two rules exist for 
creating an object inside the same \cog, or inside a newly created one.
When reaching an \key{await} statement that releases the current thread, the interrupted 
request is re-scheduled: it is inserted at any point in the request queue.\footnote{
$\textit{Schedule}(p,\many{q})$ is a function that inserts $p$ at any position inside 
$\many{q}$} This semantics ensures that the order of incoming requests is kept unchanged 
while the interrupted ones are re-schedule at any point in the future. This encodes the 
FIFO ordering of service requests discussed below.

Compared to the semantics of \absl appearing in~\cite{Giachino2015}, the following 
changes have been 
made: requests are organised in queues to enforce FIFO ordering of requests (see 
below); \emph{return} statements terminate the execution of the current method like in 
mainstream programming languages;
continuation is handled as a reserved variable, like \emph{destiny} and not as a special 
statements, this was necessary because of the new 
semantics for \emph{return}.
Our objective is to 
establish a correspondence between the configurations of \absl and the ones of the 
program translated in \multiasp. The proofs of lemmas and theorems presented in this 
section can be found in~\cite{HR-LMCSRR-2017}.

To reason on the equivalence between \absl and \multiasp configurations, we adopt the 
following conventions on object locations and activity names. We
identify activity names of \multiasp with \COG names (ranged over by $\alpha$, $\beta$). 
Also, object
locations are valid only locally in \multiasp and globally in \absl, their equivalent 
global
reference is the pair (activity name, identifier). We suppose
that each object name in \absl is of the form $i\_\alpha$ where $\alpha$ is the name of 
the \COG where
the object is created, and where $i$ is unique locally in activity $\alpha$. The 
semantics then allows us to
choose the \multiasp identifier of the created object  such that it is equal to the 
desired location ``$i$''. 

\begin{figure}[t]
	\centering
	\begin{small} 
$
		\begin{array}{rcl@{\qquad\qquad}rcl}
		\Cn & ::= & \epsilon \sep \nt{fut}(f,\nt{val}) \sep
		\nt{ob}(o,a,p,\many{q}) 
		\sep \nt{cog}(c,\nt{act})\sep 
		\Cn~\Cn & 	\nt{act}&::=& o \sep
		\varepsilon
		\\
\nt{p} & ::= & q \sep \key{idle} 

	 & \nt{val}&::=& \nt{v}\sep \bot
		\\
		q & ::= &  \{l \mid s\} 
		& a & ::= &  \vect{x\mapsto v}  
		\\
		 \nt{v} & ::= & o \sep f \sep \enull\bnfor \text{\it primitive-val} 
		\end{array}
$
	\end{small}
	\caption{Runtime syntax of \absl.}
	\label{fig:abs-runtime}
\end{figure}
\begin{figure}[t]
	\centering
	\renewcommand{\arraystretch}{0.9} 
	\begin{scriptsize}  
		$\begin{array}{@{}c@{}}
		\qquad 
		\nredrule{Skip}{
			\ob(o,a,\{l \mid \key{skip};s\},\many{q}) 
			\\
			\absred {\ob(o,a,\{l \mid s\},\many{q})}}
		\hfill
		\ntyperule{Assign-Local}{
			x \in \dom(l) \quad v=\feval{e}{(a+l)}^A
		}{
			\ob(o,a,\{l \mid x=e;s\},\many{q})
			\\
			\absred 
			\ob(o,a,\{l [x\mapsto v]\mid s\},\many{q})
		}
		\hfill
		\ntyperule{Assign-Field}{ 
			x \in \dom(a)\setminus\dom(l) \quad v=\feval{e}{(a+l)}^A
		}{
			\ob(o,a,\{l \mid x=e;s\},\many{q})
			\\
			\absred 
			\ob(o,a [x\mapsto v],\{l \mid s\},\many{q})
		}
		\qquad
		\\
		\\
		\qquad
		\ntyperule{Cond-True}{	
			\true = \feval{e}{(a+l)}^A
		}{ \ob(o,a,\{l \mid \key{if}\ e\ 
		\key{then}\:\{s_1\}\:\key{else}\:\{s_2\};s\},\many{q})\\
			\absred \ob(o,a,\{l \mid s_1;s\},\many{q})}
		\hfill\qquad
		\ntyperule{Cond-False}{ \false = \feval{e}{(a+l)}^A	
		}{ \ob(o,a,\{l \mid \key{if}\ e\ 
		\key{then}\:\{s_1\}\:\key{else}\:\{s_2\};s\},\many{q})\\
			\absred \ob(o,a,\{l \mid s_2;s\},\many{q})}
		\qquad
		\\
		\\
		\qquad \quad
		\ntyperule{Await-True}{
			f = \feval{x}{(a+l)}^A  \quad v \neq \bot
		}{
			\ob(o,a,\{l \mid \key{await} \, x \, \key{?} ; s\},\many{q})\ 
			\textit{fut}(f,v)
			\\
			\absred 
			\ob(o,a,\{l \mid s\},\many{q})\ \textit{fut}(f,v) 
		}
		\hfill
		\ntyperule{Await-False}{
			f = \feval{x}{(a+l)}^A \quad \textit{Schedule}(\{ l \mid
			\key{await}\, x \, \key{?} ;s \},\many{q})=\many{q'}
		}{
			\ob(o,a,\{l \mid \key{await} \, x \, \key{?} ; s\},\many{q})\ 
			\textit{fut}(f,\bot) 
			\\
			\absred 
			\ob(o,a,\key{idle} ,\many{q'})\ \textit{fut}(f,\bot) 
		}	
		\qquad \quad
		\\
		\\
		\nredrule{Release-Cog}{
			\ob(o,a,\key{idle},\many{q})\ \cog(c, o)
			\\	
			\absred
			\ob(o,a,\key{idle}, \many{q})\ \cog(c, \epsilon)
		}
		\quad
		\ntyperule{Activate}{
			c = a(\text{cog})
		}{
			\ob(o,a,\key{idle},\{ l \mid s\}::\many{q})\ \cog(c, \epsilon)
			\\
			\absred 
			\ob(o,a,\{ l \mid s\},\many{q})\ \cog(c, o)
		}
		\quad
		\ntyperule{Read-Fut}{
			f=\llbracket e\rrbracket_{(a+l)}^A \quad v \neq \bot
		}{
			\ob(o,a,\{l \mid x=e . \key{get};s\},\many{q})~\textit{fut}(f,v)
			\\    
			\absred 
			\ob(o,a,\{l \mid x=v;s\},\many{q})~\textit{fut}(f,v)
		}
		\\
		\\
		\qquad
		\ntyperule{New-Object}{
			o'  = \text{fresh}(\name{C}) \quad \fields{\name{C}} =  \many{x}
			\\ \many{v} = \feval{\many{e}}{(a+l)}^{A} \quad
			a'=[ \many{x \mapsto v}, \text{cog} \mapsto c]
		}{
			\ob(o,a,\{l \mid x=\key{new}\ \key{local}\ \name{C}(\many{e});s\},\many{q})\ 
			\cog(c, 
			o)
			\\
			\absred 
			\ob(o,a,\{l \mid x=o';s\},\many{q})\ \cog(c, o) ~~
			\ob(o',a',\key{idle},\emptyset)
		}
		\hfill
		\ntyperule{New-Cog-Object}{
			c' = \text{fresh}(\,) \quad o' = \text{fresh}(\name{C}) \quad 
			\fields{\name{C}} =  \many{x} \\
			\many{v} = \feval{\many{e}}{(a+l)}^{A}  \quad a'=[ \many{x \mapsto
				v}, \text{cog} \mapsto c']
		}{
			\ob(o,a,\{l \mid x=\key{new}\ \name{C}(\many{e});s\},\many{q})
			\\
			\absred 
			\ob(o,a,\{l \mid x=o';s\},\many{q})
			\\
			\ob(o',a', \key{idle}, \emptyset) \quad \cog(c', \epsilon)
		}
		\qquad
		\\
		\\
		\ntyperule{Rendez-vous-Comm}{
			f = \text{fresh}(~) \quad o'=\llbracket e\rrbracket^A_{(a+
				l)} \quad \many{v}=\llbracket\many{e}\rrbracket^A_{(a+
				l)}\quad q''=\bind(o',f,m,\many{v},\class(o'))
		}{\\[-2ex]
			\ob(o,a,\{l \mid x=e \key{!} 
			\m(\many{e});s\},\many{q})~\ob(o',a',p,\many{q'})
			\\
			\absred 
			\ob(o,a,\{l \mid x=f;s\},\many{q})~\ob(o',a',p,\many{q'}:: 
			q'')~\textit{fut}(f,\bot)
		}	
		\qquad\qquad\qquad
		\ntyperule{Context}{
			\Cn \absred \Cn'
		}{
			\Cn \; \Cn''
			\absred \Cn' \; \Cn''
		}
		\\
		\\
		\qquad \ntyperule{Cog-Sync-Call}{
			o'=\feval{e}{(a+l)}^A \quad \many{v}=\feval{\many{e}}{(a+l)}^A \quad 
			f = \text{fresh}(\,)
			\\
			c= a'(\text{cog}) \quad f'=l(\text{destiny}) 
			\\ 
			\{l' \mid s'\} = \text{bind}(o', f, \m, \many{v}, class(o'))
		}{\\[-2ex]
			\ob(o,a,\{l \mid x=e \key{.} \m(\many{e});s\},\many{q})~ 
			\ob(o',a',\key{idle},\many{q'})\ 
			\cog(c, o)\\
			\absred \ob(o,a,\key{idle},\many{q} :: \{l \mid \key{await} \; f \key{?} ; 
			x=f . 
			\key{get};s\})\ \textit{fut}(f,\bot)
			\\ \ob(o',a',\{l'[cont\mapsto f'] \mid s'\},\many{q'})\ \cog(c, o')
		}
		\hfill
		\ntyperule{Cog-Sync-Return-Sched}{
			v = \feval{e}{(a+l)}^A  \quad c = a'(\text{cog})  \\ f = l'(\text{destiny}) 
			\quad l(\key{cont}) = f
		}{
			\ob(o,a,\{l \mid \key{return}\ e;s \},\many{q})\ \cog(c, o)\\
			\ob(o',a',\key{idle},\many{q'} :: \{l' \mid s\}:: \many{q''} )\ 
			\textit{fut}(f,\bot) \\
			\absred \ob(o,a,\key{idle},\many{q})\ \cog(c, o')\\
			\ob(o',a',\{l' \mid s\},\many{q'} :: \many{q''}) \ \textit{fut}(f,v)
		} \qquad
		\\
		\\
		\ntyperule{Self-Sync-Call}{
			f'=l(\text{destiny}) \quad o=\feval{e}{(a+l)}^A \quad 
			\many{v}=\feval{\many{e}}{(a+l)}^A
			\\
			f= \text{fresh}(\,)\quad \{l' \mid s'\} = \text{bind}(o, f, \m, \many{v}, 
			class(o))
		}{
			\ob(o,a,\{l \mid x=e \key{.} \m(\many{e});s\},\many{q})\\
			\absred \ob(o,a,\{l'[cont\mapsto f'] \mid s'\},\many{q} :: \{ l \mid 
			\key{await} \; 			f \key{?} ;			x=f . \key{get};s\})\\
			\textit{fut}(f,\bot)
		}
		\qquad
		\ntyperule{Rem-Sync-Call}{
			o'\!=\!\feval{e}{(a+l)}^A \quad f= \text{fresh}(\,)
			\quad  a(\text{cog}) \neq a'(\text{cog})\\
			q''=\bind(o',f,m,\many{v},\class(o'))
		}{\\[-2ex]
			\ob(o,a,\{l \mid x=e \key{.} \m(\many{e});s\},\many{q})\ 
			\ob(o',a',p,\many{q'})\\
			\absred \ob(o,a,\{l \mid f=e \key{!} \m(\many{e}); x=f . 
			\key{get};s\},\many{q})\\ 
			\ob(o',a',p,\many{q'}::q'')\ \textit{fut}(f,\bot)} 
		\\
		\\
		\qquad
		\ntyperule{Return}{
			v = \feval{e}{(a+l)}^A  \quad f = l(\text{destiny}) \quad \nt{cont}\notin 
			\dom(l)
		}{
			\ob(o,a,\{l \mid \key{return}\ e;s\},\many{q}) \ \textit{fut}(f,\bot)
			\\
			\absred 
			\ob(o,a,\key{idle},\many{q}) \ \textit{fut}(f,v)
		}
		\qquad\qquad\qquad
		\ntyperule{Self-Sync-Return-Sched}{
			v = \feval{e}{(a+l)}^A  \quad f = l'(\text{destiny}) \quad l(\key{cont}) = f
		}{
			\ob(o,a,\{l \mid \key{return}\ e;s\},\many{q}\! ::\! \{l' \mid s\}:: 
			\many{q'})\ 
			\textit{fut}(f,\bot)\\
			\absred \ob(o,a,\{l' \mid s\},\many{q} :: \many{q'})\ \textit{fut}(f,v)
		}
		\end{array}$
	\end{scriptsize}
	\caption{Semantics of \absl.} 
	\label{fig:abs-semantics}
\end{figure}

\subsubsection{Restrictions of the translation}
The proof of correctness of the translation is valid under four specific restrictions. 
We detail them below.

\begin{enumerate}
	
	\item \textbf{Causal ordering of requests} (applies in both directions of the 
	equivalence). \multiasp ensures causal ordering of
	communications with a  rendez-vous that precedes all asynchronous method calls:  
	the request
	\emph{is dropped} in the remote request queue \emph{synchronously}. This brief
	synchronisation does not exist in \absl where requests
	can arrive in any order. The difference between  
	the modes of message 
	transmission being unrelated to our study, we constrain the semantics of \absl so 
	that it 
	uses the same communication timing as \multiasp. Thus in the semantics of 
	Figure~\ref{fig:abs-semantics}, 
	the message sending and reception rule of the original \absl\ is replaced by the 
	\textsc{Rendez-vous-Comm} 
	rule.
%
	
	\item \textbf{FIFO service of requests} (applies in both directions of the 
	equivalence). In \multiasp, while thread activation can happen in any order, the 
	order in which
	requests are served is FIFO by default instead of the non-deterministic activation of 
	a
	thread featured by \absl semantics.
	In both the Java backend  and the  \proactive backend, activation 
	and request service are FIFO, although \proactive supports the definition of 
	different policies through multi-active object 
	annotations~\cite{Henrio:2014:DSA:2554850.2554957}.  Consequently, we only reason
	on executions that enforce a FIFO policy, i.e. executions that serve requests and 
	resume
	them in a FIFO order. The semantics of Figure~\ref{fig:abs-semantics} enforces such a 
	FIFO service of incoming requests, while interleaving the service of interrupted 
	requests in a non-deterministic manner.
	
	\item \textbf{Absence of `futures of futures'} (applies in the direction from \absl 
	to \multiasp).
The semantics of futures is very different in the two languages and must be handled 
carefully. 
	 Firstly, a 
	variable holding a pointer to a 
	future object in \multiasp is equivalent to the same variable holding directly the 
	future 
	reference in \absl (several subsequent references might be followed this way). This 
	is because the \multiasp semantics use additional locations to handle transparent 
	futures. Secondly, the
	equivalence can follow future references in \absl. This is because future update 
	occurs transparently in \multiasp while in \absl they only happen upon a \get\  
	operation. Thus in the \multiasp configuration, more future updates might have been 
	performed compared to the \absl configuration. However following (future or location) 
	references is not always sufficient. Indeed, consider   the \absl configurations (i) 
	\begin{small}$\textit{fut}(f,f')~
		\textit{fut}(f',\undefined)$\end{small} and the configuration (ii) 
	\begin{small}$\textit{fut}(f,\undefined)$\end{small}; they are
	observationally
	different, whereas in \multiasp they are not because any synchronisation on the 
	future $f$ will synchronise on the future $f'$ too. Consequently we restrict  
	ourselves to \emph{programs that do not create futures of futures}, i.e. that cannot 
	create terms of the form $\textit{fut}(f,f')$ in any reachable configuration.
	The transparency of futures and of future updates creates an intrinsic an 
	unavoidable difference
	between the two active object languages. However, this is not a major restriction on 
	expressiveness because it
	is still possible to have a wrapper for futures values: a future value that is an 
	object containing
	a future. 
	Such a wrapper could be created by the translation but this would  
	break the one-to-one mapping between \absl and \multiasp futures. 
	
	\item \textbf{Forward-based distributed future updates} (applies in the direction 
	from\newline \multiasp to \absl).
	The distributed future update mechanism of \multiasp cannot be strictly faithfully 
	represented in \absl. 
	The problem arises when futures are transmitted between activities, for example when 
	a future is a parameter of a request sent to another activity. In this case in 
	\multiasp, two proxies for the same future will exist in the store of the two 
	activities, whereas only one centralised future exists in \absl. This situation can 
	create intermediate states in \multiasp where the future value is available but not 
	completely propagated to all the locations where the future has been transmitted. 
	This can create behaviours that are not possible to reach in \absl, because a future 
	update is atomically made available to all activities referencing the future. This 
	difference is only observable when the flow of method invocations races with the flow 
	of future updates.
	Enforcing an atomic future update in \multiasp would directly solve the 
	issue but this is not  realistic in a distributed setting.
	The solution we advocate for handling the potential inconsistency in future 
	resolution 
	is related to the future update mechanism of \proactive: the 
	value of a future is transmitted between activities along the same chain as the 
	future was transmitted originally. 
	We  extend the causally-ordered communications  to the 
	transmission of future values (instead of just having causally ordered request 
	sending). This prevents any observable inconsistency in the future updates and 
	simulates faithfully the \absl behaviour.
	This solution does not require  many additional synchronisations; it is reasonable 
	and relevant in a distributed 
	context.
	
\end{enumerate}

Restrictions (1) and (2) already exist in the Java backend for \absl. And  we can notice 
that these differences are more 
related to request scheduling policies and to communication channels than to the nature 
of the two active object languages. Also, restriction (4) is related to distributed 
systems, and any distributed translation would have to deal with some inconsistency in 
the observability of a future result. In the end, restriction (3) is the only one that 
gives a real insight on the differences of the two active object languages, because it is 
related to way the futures are designed and handled in the languages.
\subsubsection{Equivalence relation}

We  define an equivalence relation $\Rc$ between \multiasp and \absl terms.  This 
equivalence relation aims at proving that any single step of one calculus can be 
simulated by a sequence
of steps in the other. The notion of observational equivalence is a bit similar to the 
proof 
in~\cite{Dam:2015:LRP:2837374.2837465}. In particular, we can use the same observation 
notion: processes are observed based on remote method
invocations. The equivalence relation $\Rc$ consists of three parts: equivalence of 
values, equivalence of statements, and equivalence of configurations.
In the following, we  use the notation $\Cn$ for an \absl configuration and  
$\underline{\Cn}$ for a \multiasp configuration. 

\medskip
\begin{defi}[Equivalence of values]\label{def:translation-values}
$\approx_\sigma^{\Cn}$ is an equivalence relation between values (or between a value and a storable), in the context of
a \multiasp store $\sigma$ and of an \absl configuration $\Cn$, such that:
	\begin{small}
  	\begin{mathpar}
    \inferrule{}{v\approx_\sigma^{\Cn} v}
    
    \inferrule{}{f\approx_\sigma^{\Cn} f}
    
  \inferrule{}{i\_\alpha\approx_\sigma^{\Cn} [\nt{cog}\mapsto\alpha,\nt{Id}\mapsto i,\many{x\mapsto
        v'}]}      
\\
    \inferrule{v\approx_\sigma^{\Cn} \sigma(o)}{v\approx_\sigma^{\Cn} o}
    
    \inferrule{\textit{fut}(f,v')\in \Cn\\v'\approx_\sigma^{\Cn}v} {f\approx_\sigma^{\Cn}  v}
  	\end{mathpar}
	\end{small}
\end{defi}
	
Runtime values in \absl are either object references, future
references, or primitive values. The equivalence relation $\approx_\sigma^{\Cn}$ 
specifies 
that two values or  futures are equivalent if they are the same. An object
is only characterised by its identifier and its \COG name. The two last cases
are more interesting and reflect the difference
between the future update mechanisms of \absl and \multiasp. First, the equivalence can follow as
many local indirections in the \multiasp store as necessary. Second, the
equivalence can  follow future references in \absl, because a future can
be updated transparently in \multiasp while in \absl the explicit
future read has not occurred yet.

\begin{defi}[Equivalence of statements]\label{def:translation-statements}
%
~\\
$s\approx_{(\sigma+\ell')}^{\Cn}s'$ \iff
either $\llbracket s\rrbracket=s'$ or $s=(x=v;s_1)~ \land  ~
 	s'=(x=e;\llbracket s_1\rrbracket)~ with~ v\approx_\sigma^{\Cn}\llbracket 
 	e\rrbracket_{(\sigma+\ell')}$.
\end{defi}	
	
Two \multiasp~and \absl~statements are equivalent if one is
the translation of the other, or if both  start with an assignment of equivalent
values to the same variable, followed by an \absl statement on one side and its 
translation on the \multiasp side.

Finally, we define an equivalence between \absl\ and \multiasp\ configurations; note that 
the following definition only deals with \multiasp\ configurations that are obtained when 
evaluating the translation ob an \absl\ program.

\begin{figure}[t]
      \begin{small}
{\flushleft
$\texttt{(1a)}$        $\forall \alpha.\quad  \exists a.\,cog(\alpha,a)\in \Cn$ \iff~ 
$\exists 
o_\alpha,\sigma,p,\Rq .\,
        act(\alpha,o_\alpha,\sigma,p,\Rq)\in\underline{\Cn}$~~ \WITH ~~ $\forall i.$\\
$\texttt{(1b)}$
        \text{~}  $\exists \many{v},p',\many{q}.\ob(i\_\alpha, \manyMap{x\!\mapsto\! v}, 
        p',         \many{q})\!\in\! \Cn$ \iff  \,  $\exists o,\many{v'} 
        .\,\sigma(o)\!=\![\nt{cog}\!\mapsto\!\alpha,\nt{myId}\mapsto i, 
        \manyMap{x\!\mapsto\! v'}]$ \WITH\\
$\texttt{(1c)}$
        \text{~}\quad $
          \many{v}\approx_\sigma^{\Cn}\many{v'} ~~\land
\texttt{(1d)}    \qquad      \left(
            \begin{array}{@{}l@{}}
              \exists l,s.\, p'=\{l|s\} \iff~ \exists 
              f,i,\m,\many{v''},\ell',s',\ell'',s''.
              \left( 
              (f,\nt{execute},i,\m,\many{v''})_A\mapsto\{\ell'|s'\}::\{\ell''|s''\}\in 
              p \right. \\ \left. \land~\ell'(\ethis) = o 
              \vphantom{(f,\nt{execute},i,\m,\many{v''})}\right)
              ~\WITH~ \forall x\in dom(l)\backslash destiny.\,l(x)\approx_\sigma^{\Cn}\ell'(x)\land
              l(destiny)=f) \land s\approx_{\sigma+\ell'}^{\Cn} s'
            \end{array}\right)
          ~~\land
\texttt{(1e)}   \qquad       \forall f.\left(
            \begin{array}{l}     \big(\exists l,s.(\{l|s\}\in \many{q}\land 
            l(\nt{destiny})=f) 
            \iff \\ 
            \exists  i,\m,\many{v''},\ell',s',\ell'',s''.
              \big(\left((f,\nt{execute},i,\m,\many{v''})_P\mapsto
              \{\ell'|s'\}::\{\ell''|s''\}\in p\land \ell'(\ethis) = o\right)  \\ \lor~
              \left((f,\nt{execute},i,\m,\many{v''})\in\Rq  \land 
              o_\alpha.\nt{retrieve}(i)\!=\!o\land\bind(o,\m,\many{v''})=\{\ell'|s'\}\right)\big)\\
              ~~\WITH  \quad (\forall x\in 
              \dom(l)\backslash\{\nt{destiny}\}.\,l(x)\approx_\sigma^{\Cn}\ell'(x) ) 
              \land 
              s\approx_{\sigma+\ell'}^{\Cn}s'\big) \end{array}\right)$

\medskip
  \noindent 
$\texttt{(2)}$  $\forall f.\quad \exists v.\textit{fut}(f,v)\in \Cn$ \iff~ $\exists 
v',\sigma.\, 
(\fut(f,v',\sigma)\in \underline{\Cn}\land\Method(f)=\nt{execute})$ \WITH~
  $v\approx_\sigma^{\Cn} v'$

\medskip  \noindent 
$\texttt{(3)}$  $\forall f.\quad  \textit{fut}(f,\undefined)\in \Cn$ \iff 
~$\fut(f,\undefined)\in 
\underline{\Cn}\land\Method(f)=\nt{execute}$

}
	\end{small}
	\caption{Equivalence between \absl~and \multiasp~configurations. $\Method(f)$
  returns the method of the request corresponding to future
  $f$. We use the
  notation: $\exists y.\, P~ \iff~ \exists x.\, Q~ \WITH~ R$, meaning: $(\exists y.\, P~
  \iff~ \exists x.\, Q) \land \forall x,y.\, P\land Q\Rightarrow R$. This allows $R$ to refer to 
  $x$ and $y$.\protect\footnotemark}
	\label{fig:translation-equivalence}
\end{figure}

\medskip
\sloppy
\begin{defi}[Equivalent  configurations]\label{def:translation-configurations}
The \absl configuration $\Cn$ and the \multiasp configuration $\underline{\Cn}$ 
are equivalent,
written $\Cn~ \Rc~ \underline{\Cn}$, iff the three condition of 
Figure~\ref{fig:translation-equivalence} hold.
\end{defi}\fussy
In more details, the equivalence of Figure~\ref{fig:translation-equivalence} 
globally considers three cases:

\medskip
\noindent -- The first five lines deal with equivalence of \COGS. This case compares both
activity content and activity requests on the \absl and \multiasp sides:

\begin{itemize}
\item[\begin{scriptsize}$\bullet$\end{scriptsize}] To compare objects (Lines 
\code{1a-1c}),
we 
  rely on the fact that activities have the same name in \absl and in \multiasp.
Each \absl object $\ob$
  must correspond to one \multiasp object in the equivalent 
  activity $\alpha$. The  object equivalent to
  $ob$ must (i) be in the store, (ii) reference $\alpha$ as its
  \COG, (iii) have $i$ as identifier (the
    corresponding \absl identifier is $i\_\alpha$), and (iv) have the other fields 
  equivalent to the ones of $\ob$.
\item[\begin{scriptsize}$\bullet$\end{scriptsize}] To compare the requests,
  we  compare 
  the tasks that exist in \absl \ob\ terms
  to the tasks that exist in the corresponding \multiasp $act$ terms. We consider  two 
  cases: 
  \footnotetext{By abuse of notation and for readability, we flatten the arguments of the 
  	request and write $(f,\nt{execute},i,\m,\many{v''})$ instead of 
  	$(f,\nt{execute},(i::\m::\many{v''}))$.}
  \begin{enumerate}
  \item  Concerning the active task (Line \code{1d}): 
the single active task of \nt{ob} in \absl (in $p'$) must have exactly one 
equivalent active task in
\multiasp (in $p$).
In \multiasp, this task must have two
  elements in the current stack\footnote{The proof only deals with asynchronous 
  invocations, and thus  the stack never contains more than two 
  elements. The synchronous invocations are identical to the Java backend and 
  can be studied in this context.}: the call to the \COG (the $execute$ call) and the 
  stacked 
  call that redirected to the targeted object $o$,
  where $o$ is equivalent to \ob. In addition,  values of local variables must be
  equivalent, except  \nt{destiny}  that must correspond to the
  future of the \multiasp request. Finally, the current thread
   of the two tasks must be equivalent according to the equivalence on
  statement.
  \item   
  Concerning the inactive tasks (Line \code{1e}), two cases are possible.
 Either the task has already started and has been interrupted: it is passive and the 
 comparison is similar to the active task
    case.
 Or the task has not started yet: there must be a
    corresponding entry in the request queue $\Rq$ of the \multiasp active object 
    $\alpha$; additionally (i) computed futures must be equivalent,
    (ii)  invoked objects must be equivalent, and
    (iii) the method body built by  \emph{bind}  must be
    equivalent to the \absl task. It is important to notice that the order of the request 
    queue $\Rq$ of the \multiasp\ object is  the same as the part of the \absl\ queue 
    $\many{q}$ that corresponds to the second case, i.e. not yet started requests.
  \end{enumerate}
\end{itemize}

\noindent -- Line \texttt{2} deals with the equivalence of resolved
futures. A future value in
\multiasp refers to its local store. Two resolved futures are equivalent if their values are equivalent. In \multiasp,
only futures from \nt{execute} method calls are considered, because they represent the applicative method calls.

\noindent -- Line \texttt{3} states that the same futures must be unresolved in both 
configurations.

\medskip
Overall, the association of the equivalence of values 
(Definition~\ref{def:translation-values}), the equivalence of statements 
(Definition~\ref{def:translation-statements}), and the equivalence of \absl and \multiasp 
configurations (Definition~\ref{def:translation-configurations}), forms the global 
equivalence relation $\Rc$. We rely on equivalence $\Rc$ to prove that our  translation 
of \absl programs into \multiasp  is correct.

\subsubsection{Preliminary Lemmas}

Before stating the two theorems ensuring the correctness of the translation, 
we provide some basic properties of \absl and of the equivalence 
relation.

\begin{lemma}[Activated object in \absl]\label{lemmacog}
In an \absl configuration, if an object $\ob$ has a non-idle active request, then there 
exists a \COG in which $\ob$ is the current active object.
  \[\ob(i\_\alpha, \manyMap{x\!\mapsto\! v}, p, \many{q})\!\in\! \Cn \land p\neq 
  \key{idle} 
  \Rightarrow \cog(\alpha,i\_\alpha)\in
  \Cn\]
\end{lemma}

\begin{lemma}[Equivalence of values]\label{eval-lemma-companion}~\qquad
$v\approx_\sigma^{\Cn} v' \Rightarrow v\approx_\sigma^{\Cn} \llbracket v' \rrbracket $
\end{lemma}

\begin{lemma}[Equivalence of evaluation functions]\label{eval-lemma}
Let $\Cn$ be an \absl configuration and suppose $\Cn~ \Rc~ \underline{\Cn}$.
Let $\ob(o\_\alpha, a, \{l|s\}, \many{q})\in \Cn$. 
By definition of $\Rc$, there exists a single activity
$\act(\alpha,o,\sigma,p,\Rq)\in\underline{\Cn}$, with 
$\sigma(o)\!=\![\nt{cog}\!\mapsto\!\alpha,\nt{myId}\mapsto i, a']$
and $(f,\nt{execute},i,\m,\many{v''})_A\mapsto\{\ell'|s'\}::\{\ell''|s''\}\in p 
\land\ell'(\ethis) = o$.
For any
 \absl expression  $e$ we have:
$\llbracket e\rrbracket_{a+l}^A\approx_\sigma^{\Cn} \llbracket e\rrbracket_{\sigma+\ell'}$
\end{lemma}

The serialisation mechanism, and the renaming of local references,
are crucial points of difference between \absl and \multiasp. 
Lemma~\ref{lemma-serial-rename}  deals with these aspects. 

\begin{lemma}[Equivalence after serialisation and renaming]\label{lemma-serial-rename}
\[
\begin{array}{l}
v\approx_\sigma^{\Cn} v'\land\sigma'=\serialise(\sigma,v') \Rightarrow 
v\approx_{\sigma'}^{\Cn}v'
\\ \many{v}\approx_\sigma^{\Cn} 
\many{v'}\land(\many{v''},\sigma')=\rename{\sigma}{\many{v'},\sigma}  \Rightarrow 
\many{v}\approx_{\sigma'}^{\Cn} \many{v''}
\end{array}
\]
\end{lemma}

\noindent Besides the lemmas above, to reason on the properties of the translation,
 we rely on the fact that all \absl objects are
locally registered, in \multiasp, in the active object encoding their \COG, and we define 
an invariant for that:

\medskip
\noindent \textbf{Invariant Reg}. \emph{For every activity $\alpha$ such that
$o_\alpha$ is the location of the active object of activity $\alpha$ in its
store $\sigma_\alpha$, if the current task is an invocation to $o_\alpha.retrieve(i)$, 
then
this invocation succeeds because the object has been registered first. The invocation 
returns
some object $o'$ such that $i\_\alpha\approx_\sigma^{\Cn}o'$.}

\subsubsection{Properties of the translation}
\label{sbsc:encoding-properties}

To prove the correctness of the translation from \absl to \multiasp, we prove two 
theorems. The two theorems exactly specify under which conditions each semantics 
simulates the other. 
We first define formally the initial configuration.
 Let \begin{small}$P=\vect{I} \vect{C}\ \wrap{\vect{x}\ s}$\end{small} be an ABS
program, let \begin{small}$\Cn_0$\end{small} be the corresponding initial \absl 
configuration: \begin{small}$\ob(start, \emptyset,p,\emptyset)$\end{small}, where the 
process \begin{small}$p$\end{small} is
the activation of the program main block: \begin{small}$p=\{\manyMap{x \mapsto 
\enull}|s\}$\end{small}. 
$\llbracket 
P\rrbracket$ is the \multiasp program obtained by translation. Its 
initial
\multiasp configuration  is:
\begin{small}
	$\act(\alpha_0,o,\sigma_0[o\mapsto\emptyset],q_0\mapsto\{\manyMap{x \mapsto 
	\enull}|\llbracket
	s\rrbracket\},\emptyset).$
\end{small}
It is easy to see that this initial
configuration has the same behaviour as\\
\begin{small}
	$\act(\alpha_0,o,\sigma_0[o\mapsto\emptyset,\nt{start}\mapsto\emptyset],$ 
	$(f,\nt{execute},\nt{start},\m)\mapsto\{\manyMap{x \mapsto \enull}|\llbracket 
	s\rrbracket\}::\{\emptyset|x=\bullet\},\emptyset)$.
\end{small}

We denote this new \multiasp initial configuration 
\begin{small}$\underline{\Cn_0}$\end{small} and notice that 
$\Cn_0\Rc\underline{\Cn_0}$. 

\begin{theorem}[\absl to \multiasp]\label{thm:abs-to-multiasp} The translation simulates 
all  \absl executions
  with FIFO policy and rendez-vous communications, provided that no future value is  
  a reference to another future. 
 \[
\Cn_0\!\absredN \!\Cn \land\nexists f,f'.
\,\textit{fut}(f,f')\!\in\!\Cn~\Rightarrow~\exists
\underline{\Cn}.\,\underline{\Cn_0}\!\redN\!\underline{\Cn}~
\land~\Cn\,\Rc\,\underline{\Cn}
\]\end{theorem}
\begin{theorem}[\multiasp to \absl]\label{thm:multiasp-to-abs}  Any reduction of the 
	\multiasp translation corresponds to a  valid \absl execution.
	\[\underline{\Cn_0}\redN \underline{\Cn} \Rightarrow\exists \Cn.\,\Cn_0\absredN \Cn 
	\land 
	\Cn\,\Rc\,\underline{\Cn}\]
\end{theorem}

The proof of the first theorem can be found in~\cite{HR-LMCSRR-2017},
this proof is a classical case 
analysis on the \absl reduction rule used for evaluating the current configuration (we do 
an induction on the reduction).
Table~\ref{tab:recap-proof-th1-body} summarises the most informative part of the proof of 
Theorem~\ref{thm:abs-to-multiasp}. It shows  which \multiasp rule (second column) is used 
to simulate 
which \absl rule (first column.
The third column shows additional non-observable
steps that are introduced in the \multiasp translation. These additional steps are needed 
to reach an equivalent configuration but they
are always local and they never introduce concurrency. They typically deal with object 
registration, access to intermediate objects, or scheduling inside the \cog\ object. As 
expected, the 
proof case for the \textsc{Read-Fut} rule uses the fact that ``futures of futures'' are 
forbidden. Lemmas~\ref{eval-lemma} and~\ref{lemma-serial-rename} are used frequently to 
ensure the equivalence of the configurations, and in particular the equivalence between 
a  direct object reference of \absl and the
local \multiasp object representing an object registered in another \cog.

A detailed proof sketch for Theorem~\ref{thm:multiasp-to-abs} can also be found 
in~\cite{HR-LMCSRR-2017}. 
In this proof the
equivalence relation $\Rc$ has to be adapted because of the existence of additional 
(deterministic) statements in the translation. Table~\ref{tab:recap-proof-th2-body} 
summarises this second proof by showing
which \absl rule is used to simulate which \multiasp rule. Again, in some cases, a few 
(non-observed) additional \absl\ rules are applied to reach an equivalent 
configuration.

\begin{table}[!tp]
	\centering
	\begin{tabular}{@{}|@{}c@{}|c|c|@{}}
		\hline
		\textbf{\absl~rule} & \textbf{\multiasp~rule} & \textbf{Additional 
			\multiasp~rules} \\
		\hline
		\hline
		\textsc{Assign-Local} & \textsc{Assign-Local} & -- \\
		\hline
		\textsc{Assign-Field} & \textsc{Assign-Field} & -- \\
		\hline
		\textsc{Await-True} & -- & \textsc{Update} / -- , \textsc{Invk-Passive},\\
		&		   &  \textsc{Return-Local},
		\textsc{Assign-Local-Tmp}\\
		\hline
		\textsc{Await-False} & \textsc{Invk-Future} & -- \\
		\hline
		\textsc{Release-Cog} & -- & --\\
		\hline
		\textsc{Activate} & -- & \textsc{Activate-Thread} / (\textsc{Serve}, 
		\textsc{Invk-Passive},\\
		&		& \textsc{Return-Local},
		\textsc{Assign-Local-Tmp})\\
		\hline
		\textsc{Read-Fut} & -- & \textsc{Set-Hard-Limit}, \textsc{Update} / -- , \\
		&		& \textsc{Invk-Passive}, \textsc{Return-Local}, \\
		&		& \textsc{Set-Soft-Limit},  \textsc{Assign-Local-Tmp}\\
		\hline
		\textsc{New-Object} & \textsc{New-Object} & \textsc{Invk-Passive},	  
		\textsc{Assign-Local-Tmp} \\
		&& \textsc{Return-Local}\\
		\hline
		\textsc{New-Cog-Object} & \textsc{New-Object} & 
		\textsc{New-Active},\textsc{Assign-Local-Tmp}, \\
		&                                             & \textsc{Invk-Active-Meta}, 
		\textsc{Return} \\
		\hline
		\textsc{Rendez-vous-Comm} & \textsc{Invk-Active} & \textsc{Invk-Passive},
		\textsc{Return-Passive},\\ && \textsc{Assign-Local-Tmp} \\
		\hline
		\textsc{Return} & \textsc{Return} & \textsc{Return-Local}, 
		\textsc{Assign-Local-Tmp}\\
		\hline
	\end{tabular}
	\caption{Summary  of the simulation of \absl in \multiasp. 
		\textsc{Assign-Local-Tmp} is 
		\textsc{Assign-Local} on a variable introduced by the translation.  
		\textsc{Invk-Active-Meta}  
		is  
		\textsc{Invk-Active}  on a method that is not  $\nt{execute}$.}
	\label{tab:recap-proof-th1-body}
\end{table}
\begin{table}[!tp]
	\begin{center}\vspace{-2.6ex}
		\begin{tabular}{|c|c|c|}
			\hline
			\textbf{\multiasp~rule} & \textbf{\absl~rule} & \textbf{Additional 
				\absl~rules} \\
			\hline
			\hline
			\textsc{Assign-Local} & \textsc{Assign-Local} & -- \\
			\hline
			\textsc{Assign-Field} & \textsc{Assign-Field} & -- \\
			\hline
			\textsc{Invk-Future} & \textsc{Await-False} & \textsc{Release-Cog} \\
			\hline
			\textsc{Invk-Passive} & -- &  --/\textsc{Read-Fut}/\textsc{Await-True}\\
			\hline
			\textsc{Activate-Thread} & \textsc{Activate} & --\\
			\hline
			\textsc{Serve} & \textsc{Activate} & --\\
			\hline
			\textsc{Invk-Active} &\textsc{Rendez-vous-Comm} &--  \\
			\hline
			\textsc{New-Object} & \textsc{New-Cog-Object} & -- \\[-.4ex]
			{\small in an activity with no \absl~object}&&\\
			\hline
			\textsc{New-Object} & \textsc{New-Object} & -- \\[-.4ex]
			{\small in a non-empty activity}&&\\
			\hline
			\textsc{Return} & \textsc{Return} & -- \\
			\hline
			Others& -- & -- \\
			\hline
		\end{tabular}
	\end{center}
	\caption{Summary table of the simulation of \multiasp~in \absl.}
	\label{tab:recap-proof-th2-body}
\end{table}

\subsection{Discussion on the Translation and its 
Properties}\label{sbsc:encoding-conclusion}

In the translation, \absl requests, \COGS, and futures respectively match
\multiasp requests, active objects, and futures. For each \absl object there 
exist several copies of this object in \multiasp; all copies share the same value for the 
field \COG and identifier, but only the copy that is hosted in the 
right \COG/activity is equivalent to the \absl object. This forms a shallow translation: 
applicative requests, 
futures, active objects, and object fields are mapped faithfully by the same notion in 
\multiasp as in \absl. The main difference is that all requests transit by the \COG 
active object that acts as a local scheduler; ``execute'' 
requests  have no counterpart in \absl.
In the equivalence relation, the objects are identified  by their identifier and their \COG name, and the equivalence can follow futures. 
The equivalence between requests distinguishes two cases. First, for active tasks, there is a 
single active task per \COG in \absl and it must correspond to the single active thread 
serving an \emph{execute} request in \multiasp. Second, inactive tasks in \absl 
correspond either to passive requests being 
currently interrupted or to requests that have not been served yet in \multiasp. For each 
request that has started its execution, the second element in the stack of method calls 
corresponds to 
the invoked request, and the equivalence of 
executed statements, of local variables, and of the corresponding future is verified.

In both directions, we prove a weak simulation relation with additional non-observable 
steps dealing
with the intermediate structures created by the translation. Additional steps  also 
ensure equivalence concerning future updates. 
However, our results are  stronger than the standard guarantees given by a 
weak 
simulation because  the added steps   do not
introduce concurrency, the silent actions are always confluent.
Overall, most of the properties provided by ABS tools like absence of deadlock or 
resource consumption are guaranteed to be preserved by our translation.
%
The most striking example of an observable reduction in \absl that is not observable in
\multiasp is the update of a future with its computed value.
Indeed, the transparent creation and update of futures creates an intrinsic difference
between the two programming languages. This is why, in the first  theorem, we 
exclude the possibility for a future value to be itself a future.
Also, concerning assignments, only the ones
concerning \absl local 
variables can be 
observed in \multiasp,
assignments of temporary variables introduced by the translation have no counterpart in 
\absl. Concerning asynchronous 
method invocations, only  the applicative \absl requests are observable.

Identifying the differences of observability between active object languages  gives a 
significant insight on their  design and differences. We could 
observe the crucial differences between the language because we chose a faithful 
translation that matches most of
the elements of \absl and \multiasp configurations in a 
one-to-one way. The divergent notions of the two languages can be 
spotted  easily thanks to our shallow translation.

\subsection{Related Works and Discussion on the Execution of Active Object Programs}
\label{sbsc:encoding-related}

Using a backend allows to decouple the language for
specifying and verifying the program from the execution language, and makes it possible 
to explore easily new programming 
paradigms. Such a 
design enables writing programs and proving
properties on a high-level language, and relying on massively used
platforms for implementation. 

A closely related ongoing work~\cite{CPE:CPE3480} aims at implementing the \absl 
semantics with the paradigms of Java 8, using
a lightweight thread continuation mechanism (representative of cooperative scheduling). 
This work makes a particular focus on efficiency, as opposed to the seminal Java backend 
for \absl. Preliminary results 
showed that the Java 8 backend for \absl  scales much better than the existing Java 
backendconcerning the number of threads 
co-allocated on the same machine. However, to the best of our knowledge, this backend is 
still under development and does not support distributed execution.

In parallel with this work, ABS has been extended with \emph{deployment
	components}~\cite{Bezirgiannis2016} which are specific objects representing the
locations where new \COGS must be deployed. 
Deployment components are specific objects of the
ABS language. They are used to reason on the deployment of the application: they
represent an abstraction of the deployment location and enable the reasoning on the
distributed nature of the application. Instead we   use \emph{virtual nodes} expressing 
deployment in the \absl syntax similarly to the \proactive approach. 	Those two contrary
design decisions reveal the difference of nature between the two languages concerning
deployment. The distribution system of \absl is meant to reason on object distribution.
On the contrary, \proactive virtual nodes are meant to be used in conjunction with
external deployment tools. \proactive virtual nodes  are binders to entries in a
deployment descriptor file. The expressiveness in terms of distribution of \proactive is
closer to  languages like Akka where deployment locations are strings. Deployment
components can be manipulated and stored like any other object in \absl and cannot be
translated simply into \proactive virtual nodes that are only strings pointing to a
specific entry in a deployment file. It would be possible to use some specific 
\multiasp\  active objects to represent
deployment components, but the relation with the normal \cog\ objects would need to be
specified both in practice, involving more development, and from a theoretical point of
view, a new equivalence relation would be needed.

The Haskell backend~\cite{Bezirgiannis2016} for \absl, also focuses on a distributed 
execution of \absl programs. As opposed to Java-based backends, 
lightweight thread continuations are natively available in Haskell, which makes the 
Haskell backend for \absl very efficient even with a high degree of local parallelism: 
much more \COGS can be hosted on the same machine than with programs generated by the 
\proactive backend. In this work, the \absl compiler is extended to integrate the notion 
of deployment components within the \absl language. The Haskell backend for \absl also 
has support for garbage collection of distributed objects, built on top of the Haskell 
garbage collector. In \proactive, activation of distributed garbage collection is 
optional, but it is less crucial since all passive objects are garbage collected by the 
JVM. Nevertheless, to the best of our knowledge, the \proactive 
backend  is the only backend that generates a distributed and executable code 
formally proven to be correct with respect to the \absl semantics.

Our backend suffers from several performance limitations inherent to the approach 
and to the conceptual differences between the languages. First, in some cases, many 
threads might be blocked in a 
wait-by-necessity state and this could be improved in several ways. 
Currently, 
the \proactive library  reuses a blocked thread to serve a new request if it is 
sure that the  re-used thread cannot be re-activated while the new request is 
served: a request  re-uses a blocked thread if this request will resolve the future on 
which the thread is blocked. In other cases, the 
nature of Java (non-preemptive  with heavy threads) prevents us from doing further 
optimisations without relying on a heavier 
compilation phase, e.g. using 
continuations, and without breaking the direct correspondence between 
the two languages, and the 
simplicity of the object language translation. Another approach could consist in creating
more active objects, i.e. one per ABS object, but, as discussed in the introduction of 
this section, 
this would make the synchronisation between the different objects  tricky. 
Finally, the \await\ statement on a boolean expression relies on a busy-wait polling 
which 
can be highly inefficient, a wait-notify translation could be more efficient provided the 
monitoring of field mutation is done efficiently~\cite{AzadbakhtBB-sofsem17}. 
Wait-notify is already used to implement wait-by-necessity in \proactive, it is thus used 
by \await\ operations on futures. We however do not use wait-notify for the 
computation of boolean expressions because this would require to explicitly do a wait 
(upon evaluation of the expression) and a notify (upon modification of the related 
variables). 
Wait and notify are not ASP primitives and do not follow the active-object paradigm, we 
cannot use them in the current approach.

These limitations illustrate the different approaches of the two languages: \proactive 
library was designed to compose distributed applications made of heavy threads with 
relatively few synchronisation points. \absl was designed to model parallel applications 
with many synchronisations and no problem of distribution. \proactive shines for 
implementing distributed 
applications typical of high-performance computing: intensive computations that feature 
complex synchronisations but not many threads and synchronisation points per machine.
Consequently our backend is very 
efficient for applications that do not do many cooperative thread-release or many 
conditional awaits. Indeed, we showed that such HPC applications can be run 
efficiently with the \proactive backend.


\section{Conclusion}
\label{sec:conclusion}

This paper presents a  framework for multi-active  
objects, featuring programmer-friendly concurrent programming, advanced request 
scheduling mechanisms, and development support. We  give the guidelines of the 
programming model usage through practical  applications.  The implementation 
of multi-active objects given by \proactive offers a set of annotations that 
allow the programmer to control the  multi-threaded execution of active 
objects. The formalisation of the \proactive library in the 
\multiasp programming language  provides an operational 
semantics to reason on multi-active object executions.
Priorities and thread management, combined with  multi-active object 
compatibilities, reveal to be convenient to encode scheduling patterns. This article 
 presents the \proactive backend for 
\absl. This backend  automatically transforms 
\absl models into distributed applications. We  formalise the translation and 
 prove its correctness. We establish an equivalence 
relation between \absl and \multiasp configurations, and we prove two theorems 
that corroborate the correctness of the translation, illustrating the 
conceptual differences between the two languages.

Overall, we attach a particular interest to three objectives: usability,  
correctness, and performance of our framework. We address the complete spectrum 
of the multi-active object programming model, from design to execution. Testing 
our developments in realistic settings assesses the efficiency of our 
implementation.  On the other hand, our work is formalised and we highlight the 
properties of our model. Thus, we  believe that we reinforce the guarantees offered 
by the multi-active object programming model, on which the programmer can 
 rely. 
 
In the future, we want to investigate the verification of compatibility annotations, but 
also the dynamic tuning of thread management aspects where the thread scheduler could 
adjust at runtime  the number of threads allocated depending on the 
status of the machine and of the active object.

\section*{Acknowledgement}
Experiments presented in this paper were carried out using the Grid'5000 testbed, 
supported by a scientific interest group hosted by Inria and including CNRS, RENATER and 
several Universities as well as other organizations (see https://www.grid5000.fr).
We would like to thank the reviewers of this article for their thorough work and 
constructive comments.

\bibliographystyle{abbrv}
\bibliography{biblio}
%
%

\end{document}